\documentclass{my}
\begin{document}
\newcommand{\average}[1]{\mbox{$ \langle #1 \rangle $}}
\newcommand{\mean}[1]{\left< #1 \right>}
\newcommand{\lambdams}{\mbox{$\Lambda_{\rm \overline{MS}}$}}
%                                                    units
\newcommand{\lsim}{\raisebox{-1.5mm}{$\:\stackrel{\textstyle{<}}{\textstyle{\sim}}\:$}}
\newcommand{\gsim}{\raisebox{-0.5mm}{$\stackrel{>}{\scriptstyle{\sim}}$}}
\newcommand{\pom}{\rm I\!P}
\newcommand{\reg}{\rm I\!R}
\newcommand{\alphapom}{\alpha_{_{\rm I\!P}}}
\newcommand{\alphareg}{\alpha_{_{\rm I\!R}}}
\newcommand{\xpom}{x_{_{I\!\!P}}}
\newcommand{\mx}{M_{_X}}
\newcommand{\my}{M_{_Y}}
\newcommand{\gapprox}{\stackrel{>}{_{\sim}}}
\newcommand{\lapprox}{\stackrel{<}{_{\sim}}}
\newcommand{\etamax}{\eta_{\rm max}}
\newcommand{\zpom}{z_{_{\rm \pom}}}
\newcommand{\xgam}{x_{\gamma}}
\newcommand{\zpomj}{z_{_{\rm \pom}}^{\rm jets}}
\newcommand{\xgamj}{x_{\gamma}^{\rm jets}}
\newcommand{\av}[1]{\mbox{$ \langle #1 \rangle $}}
\title{\vspace{-0.5cm} Small-x QCD Effects in Particle Collisions at High Energies}
\author{\vspace{-0.4cm} Tancredi Carli}
\address{DESY, Notkestr. 85, Hamburg
Germany, E-mail: carli@mail.desy.de \\
Invited talk at XX Int. Symp. on Lepton and Photon Interactions at High Energies,\\
 Rome, Italy, 23-28 July 2001.
 \vspace{-0.4cm}} 
\twocolumn[\maketitle\abstract{
Recent theoretical developments to calculate cross sections 
of hadronic objects in the high energy limit
are summarised and experimental
attempts to establish the need for new QCD effects 
connected with a resummation of small hadron momentum fractions $x$
are reviewed. 
%beyond the standard DGLAP approach 
The relation between
small-$x$ parton dynamics and the phenomenon of diffraction 
is briefly out-lined. In addition, a search for
a novel, non-perturbative QCD effect, the production of
QCD instanton induced events, is presented.
}]
%%%%%%%%%%%%%%%%%%%%%%%%%%%%%%%%%%%%%%%%%%%%%%%%%%%%%%%%%%%%%
\section{Mini-Introduction to Recent BFKL Progress}
\label{subsec:intro}
When two hadronic objects collide at high energies,
$h_1 \, h_2 \to h_3 \, h_4 \, X$, often more than one energy scale
%(larger than the QCD scale $\Lambda_{QCD}$)
is involved. Of particular interest are collisions
where the squared centre of mass (cm) energy $s = (h_1 + h_2)^2$
is much larger than the squared transverse 
momentum transfer
$t = (h_1 - h_3)^2$, i.e. $s \gg t \gg \Lambda_{QCD}$. 
In perturbative QCD (pQCD) 
calculations of such processes large logarithms of the
ratio of these scales, $(\alpha_s \ln{(s/t)})^n$, 
arise at each order $n$ in the strong coupling constant
$\alpha_s$. It is usually assumed that they must be resummed to obtain a reliable
cross section prediction.

About 25 years ago, Balitsky, Fadin, Kuraev and Lipatov
(BFKL) derived in the leading logarithmic approximation (LLA)
an equation describing the resummation of a certain class of exchanges
with one chain of multiple gluons in the $t-$channel\cite{bfkl}. It
predicts a power-law rise of the scattering
cross section at high energies: 
%%%%%%%%%%%%%%%%%%%%%%%%%%%%%%%%%%
\vspace{-0.2cm}
\begin{eqnarray}
\sigma_{\footnotesize BFKL}^{\footnotesize LO} \propto s^\lambda 
 \;  {\rm with }  \; 
% \lambda =      (\alpha_s}/\pi) \, 12 \, \ln{2} \;\;
\lambda = \frac{\alpha_s}{\pi} \, 12 \, \ln{2} \;\;
%\nonumber
\label{eq:lambfkl} 
\vspace{-0.3cm}
\end{eqnarray}
%%%%%%%%%%%%%%%%%%%%%%%%%%%%%%%%%%%%
where $\alpha_s$ is taken at the value of a hard scale involved
in the process.
Since for typical values of $\alpha_s$, e.g. $\alpha_s = 0.2$, 
$\lambda \approx 0.5$,
the predicted rise of the cross section is so strong
that the unitarity bound will be violated.
This is an indication that at high energies
novel effects like parton recombination or multiple
perturbative scatterings taming down the rise
should be important.
%%%%%%%%%%%%%%%%%%%%%%%%%%%%%%%%%%%%%%%%%%%%%%%%%%
\begin{figure}
\vspace{-0.6cm}
\epsfxsize100pt
\figurebox{}{}{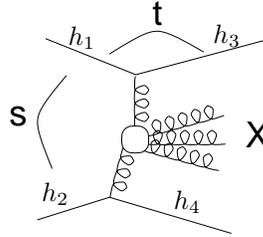}
\begin{picture}(50.,50.)
\put(-100.,72) {$h_1$}
\put(-110.,13) {$h_2$}
\put(-45.,73.) {$h_3$}
\put(-60.,10.) {$h_4$}
\end{picture}
\vspace{-0.2cm}
\caption{Sketch of a high energy collision of two hadronic objects
$h_1 \, h_2 \to h_3 \, h_4 \, X$.
}
\label{fig:ladder}
\vspace{-0.8cm}
\end{figure}
%%%%%%%%%%%%%%%%%%%%%%%%%%%%%%%%%%%%%%%%%%%%%%%%%%%%
%
Over the past years much experimental effort has been
devoted to observe to such phenomena.
Although some hints for a rise of the scattering cross section 
at high energies have been found,
the predicted large value of $\lambda$ could not be confirmed.

\begin{figure}
\epsfxsize180pt
\figurebox{}{}{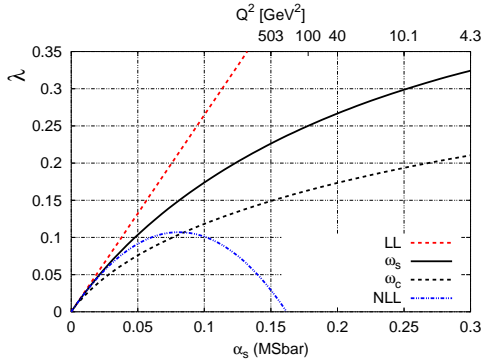}
%\vspace{0.3cm}
\caption{The BFKL exponent $\lambda$ as a function of $\alpha_s$
in the LO and NLO approximation and the collinearly
enhanced resummation included at NLO.% \cite{bfklresum}.
}
\label{fig:omcrit}
\vspace{-0.7cm}
\end{figure}

After a huge theoretical effort by many authors, 
the calculations of the next-to-leading order
(NLO) corrections to the BFKL equations 
were finalised in 1998\cite{bfklnlo}.
The NLO terms are suppressed by a power of $\alpha_s$: 
$\alpha_s (\alpha_s \ln{(s/t)})^n$. 
To the consternation of the community, the NLO corrections
turned out to be large and lead to much smaller values of
$\lambda$ than the leading order (LO) contribution. 
Fig.~\ref{fig:omcrit} shows $\lambda$ calculated in LO and NLO
as a function of $\alpha_s$.
For instance, for a scale
where $\alpha_s \approx 0.16$ the NLO
corrections exactly cancel the LO term. For larger values,
$\lambda$ even becomes negative. Naively, this would lead
to a cross section that decreases rather than increases
as a power of $s$. This is in contradiction
to the data.
The large NLO corrections put the applicability of the
BFKL resummation in question. It seems necessary to 
understand the physical reason for the large corrections
and to include them at all orders.

%%%%%%%%%%%%%%%%%%%%%%%%%%%%%%%%%%%%%%%%%%%%%%%%%%
%%%%%%%%%%%%%%%%%%%%%%%%%%%%%%%%%%%%%%%%%%%%%%%%%%%%

Several solutions 
have been proposed to solve this problem\cite{bfklsol,f2bfkl},
all of which are related to modelling and to resumming 
the higher order contributions. One promising method
is to analyse the structure of the divergences of the
BFKL characteristic function\footnote{
The characteristic function is the Mellin transform
of the BFKL kernel. The anomalous dimension is the Mellin
transform of the splitting function.
See ref.\cite{bfklintro}
for an introduction to more details.} 
at all orders by studying the limit where partons are emitted 
collinearly\cite{bfklresum}.
The collinearly-enhanced contributions have been shown to
be able to estimate the NLO corrections and there are
good reasons to believe that they also might give a significant
contribution of the higher order corrections beyond NLO.
The exponents obtained from this collinear resummation are shown 
in Fig.~\ref{fig:omcrit}.
The label $\omega_s$ is related to the power of
the cross section expected for $\gamma^* \gamma^*$ collisions or
jet observables in collisions of hadronic objects.
The line labelled $\omega_c$ is the leading pole in the
anomalous dimension of the gluon.
The difference between the two exponents
does not reflect an additional uncertainty, but is
caused by additional corrections for quantities involving 
an effective cut-off on the lowest accessible momentum 
necessary to avoid diffusion into the infra-red and non-perturbative region
and is closely related to differences in running coupling effects.
%
%Differences in the
%exponents are expected for various processes.
%
The resummation in the collinear limits
leads to values of $\lambda$ which are in qualitative
agreement with the experimental observations (see later).
However, a quantitative analysis is still lacking.

Although most of the theoretical effort was devoted to 
understanding the BFKL behaviour at NLO, most of the
phenomenological studies still use LO.
For instance, only recently part of the 
virtual photon impact factor, 
i.e. the coupling of the photon to the gluon chain, 
has been calculated in NLO\cite{impact}.

%%%%%%%%%%%%%%%%%%%%%%%%%%%%%%%%%%%%%%%%%%%%%%%%%%%%%%%%%%%
\begin{figure}
\vspace{-0.4cm}
\epsfxsize100pt
%           height width name
%\vspace{-2.cm}
\begin{tabular}{cc}
\epsfxsize90pt
\figurebox{}{20}{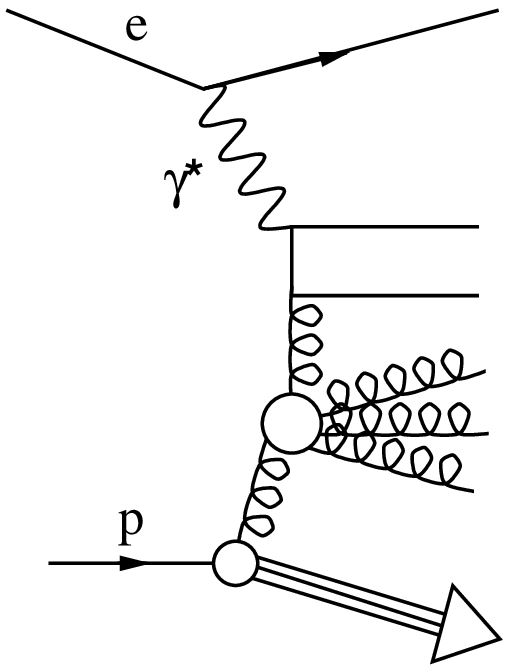}
\epsfxsize100pt 
\figurebox{}{20}{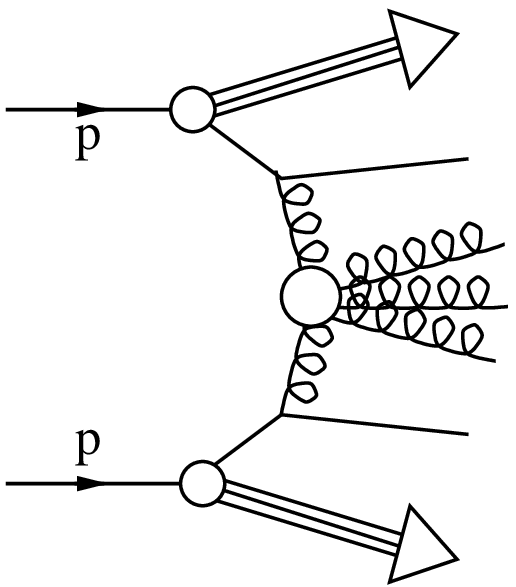} \\
\epsfxsize70pt
\figurebox{}{20}{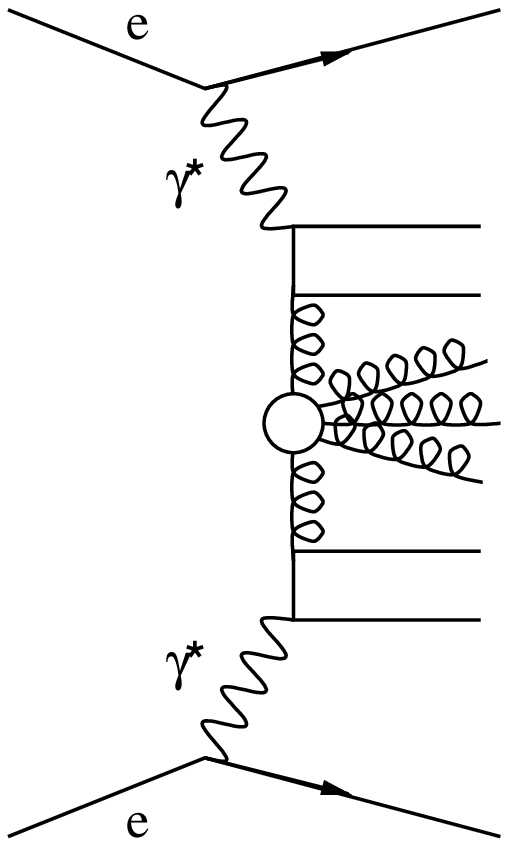}
\epsfxsize85pt
\figurebox{}{20}{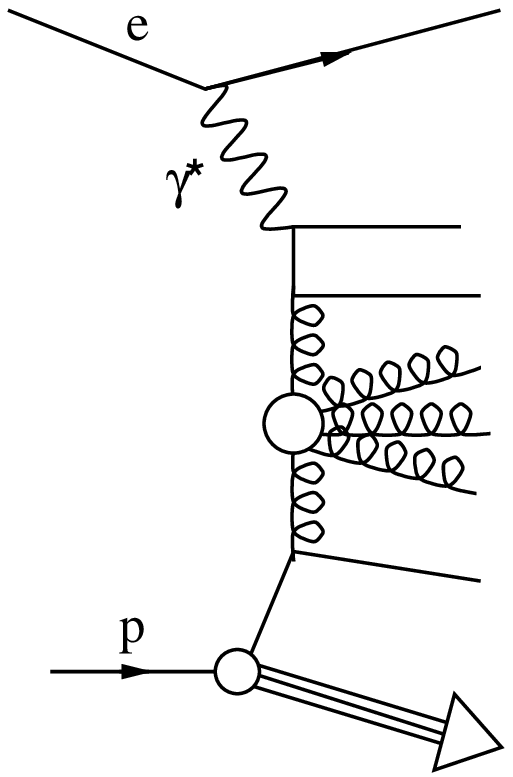} 
\end{tabular}
\begin{picture}(50.,50.)
\put(0.,50.) {(c)}
\put(0.,190.) {(a)}
\put(100.,50.) {(d)}
\put(100.,190.) {(b)}
\end{picture}
\vspace{-1.6cm}
\caption{Sketch of small-$x$ signatures in high energy 
collisions: 
a) total cross section in DIS,
b) dijet production in $p \bar{p}$ collisions,
c) total hadronic cross section in $\gamma^* \gamma^* $ collisions
d) forward jet production in DIS.
}
\label{fig:lowxsig}
\vspace{-0.8cm}
\end{figure}
%%%%%%%%%%%%%%%%%%%%%%%%%%%%%%%%%%%%%%%%%%%%%%%%%%%%
\section{Parton Evolution and DIS}
\label{sec:evolution}
In deep-inelastic scattering 
(DIS)\footnote{For an introduction 
and literature on DIS, DGLAP,
small-$x$ and related topics see: $ \;\;\;\;$\\ 
http://sesam.desy.de/$\sim$ carli/hera.html},
a photon with virtuality $Q^2 \gg m_p^2$, where $m_p$
is the proton mass, 
%much larger than the squared proton mass ($m_p^2$) 
collides with a proton.
When the squared $\gamma^* p$ cm energy $W^2$
is much larger than $Q^2$, the momentum fraction of the
struck quark, $x \approx Q^2/W^2$, becomes small. 
In this case, the simple picture of DIS
as a process where a virtual photon interacts
instantaneously with a point-like parton quark  
freely moving in the proton has to be modified.
Indeed, the probability that a gluon radiates becomes
increasingly high at small-$x$ and therefore 
the quark struck by the photon
originates from a cascade initiated
by a parton with high longitudinal momentum.    
%The phase space for gluon radiation
%is enlarged with decreasing $x$. 
A sketch of such an interaction is shown
in Fig.~\ref{fig:lowxsig}a). 
The gluon is the driving force behind the cascade
indicated by the circle.

There are three equations describing the parton evolution 
in different kinematic regions:
DGLAP, BFKL and CCFM equations.
%\vspace{-0.3cm}
\subsection{DGLAP and DLL Evolution}
To correctly calculate the inclusive DIS cross section
(for all $x$-values),
terms of the form ${(\alpha_s \ln{(Q^2/Q^2_0)})}^n$ have to be resummed,
since the smallness of $\alpha_s$ is compensated by the large
size of $\ln{Q^2}$. % ($Q^2 \gg m_p^2 > \Lambda_{QCD}^2$).
This resummation can be achieved by the DGLAP equations\cite{dglap}.
They are related to the renormalisation group equation and
describe the change of the parton density in the
proton with varying spatial resolutions $Q^2$. 
Once the parton distribution
is specified at a given scale $Q^2_0$, it can be
predicted for any other $Q^2$. 
In an appropriate gauge, the subsequent parton emissions 
from the proton to the struck quark 
(connecting the soft proton constituents to the hard subprocess)
can be interpreted as a ladder diagram whose rungs 
and emitted partons are strongly 
ordered\footnote{$k_T$ ordering is an assumption
necessary to derive the DGLAP equations.
For an introduction see ref.\cite{martin93}.}
in transverse momentum $k_T$.
This leads to a suppression of the available
phase space for gluon radiation. 

%The collinear singularities
%are factorised in the parton density functions.
%$ \sigma =  \sigma_0 \int \frac{dz}{z} C^a(\frac{x}{z}) f_a(z,Q^2)$

For high energy scattering, i.e. at small-$x$, the ladder becomes
long and consists dominantly of gluons\footnote{The splitting
function $P_{gg}(z)$ 
(describing the radiation of gluons from gluons) 
dominates at small longitudinal momentum fractions
$z$ ($P_{gg}(z) \sim 1/z$ while $P_{qq}(z) \sim$ constant).
$z$ ranges between $x$ and $1$.}.
The correct description of the small-$x$ region
is still a matter of debate and subject to many 
technical difficulties.
The dominant terms  are of the form
${(\alpha_s \ln{(1/x)} \ln{(Q^2/Q^2_0)})}^n$. Their resummation 
is referred to as the double-leading logarithmic approximation (DLL).
Both longitudinal and transverse momenta are strongly ordered.
The DLL  leads to a gluon density 
strongly rising with an effective power 
$\lambda \approx \sqrt{(12 \alpha_s/\pi) \, \ln{(1/x)} \, \ln{(Q^2/Q^2_0)}}$
towards low-$x$\cite{rujula}.
$\lambda$ depends on $x$ and $Q^2$, e.g. it increases
towards higher $Q^2$.
%
%$g(x,Q^2) \sim \exp{(2 \sqrt{\alpha_s \, \ln{1/x} \, \ln{Q^2/Q^2_0}})}$.
%                  1   2            3   3    3         321
\vspace{-0.3cm}
\subsection{BFKL Evolution}
For $Q^2 \approx Q^2_0$, double logarithms no longer
dominate and the terms 
$(\alpha_s \, \ln{(1/x)})^n \approx (\alpha_s \, \ln{(W^2/Q^2)})^n$ 
are important and must be
resummed using the BFKL equation. 
In a physical gauge, these terms correspond
to an $n$-rung ladder diagram in which gluon emissions
are ordered in longitudinal momentum, $z$. The strong $k_T$ ordering
is replaced by a diffusion pattern as one proceeds along the 
gluon chain.
The BFKL equations describe how a
particular high momentum parton in the proton is dressed by
a cloud of gluons at small-$x$ localised in a fixed transverse 
spatial region of the proton given by $Q^2$.

A %natural 
framework to describe the inclusive
DIS cross section at small-$x$ is 
%
%$ \sigma =  \int \frac{dz}{z} d^2 k_T \, C(\frac{x}{z},k_T) \, f(z,k_T,\mu^2)$
%
a $k_T$ dependent, unintegrated, gluon density $f(z,k_T)$:
\begin{eqnarray}
\sigma \propto 
\int \frac{d z}{z} \, d^2 k_T \; \hat{\sigma}(x/z,k_T) \; f(z,k_T)
\label{eq:unintgluon}
\nonumber
\end{eqnarray}
where $\hat{\sigma}(x/z,k_T)$ is the partonic cross section.
%$x g(x,\mu^2) = \int_0^{\mu^2} (d^2 k_T/k_T^2) \; x f(x,k_T^2,\mu^2)$.
% 
Together with the $k_T$ factorisation\cite{ktfac}
this forms a basis for a general calculation scheme
for small-$x$ processes. 
%The hard scale $\mu^2$
%acts as factorisation scale and controls the ordering
%of the emitted partons.

At very small-$x$, it is expected that many gluons coexist and will
no longer act as free partons, but interact with each other. 
This ``saturation'' regime is characterised by an equilibrium 
of gluon emission and absorption.
%A new QCD regime is reached where individual parton-parton 
%interactions are weak, but where the field strength becomes
%- due to the number of partons - so strong that perturbation 
%theory is not reliable.
In the regime accessible for HERA such effects are expected to be small\cite{kimber},
but could be rather large when lower
$x$-values could be reached. At THERA colliding the $500$~{\rm GeV} electrons
from TESLA with the $920$~{\rm GeV} proton from HERA such effects will be visible.
Also if HERA collided heavy ions instead of protons, an experimental
study of saturation effects would probably be possible.

%\footnote{Before the start-up of HERA
%in some theories, which assumed that there are small, 
%localised regions of 
%enhanced probability to find partons in the proton 
%(so-called ``hot spots''), the observation of such
%effects at HERA was considered to be possible\cite{hotspotsXX}. 
%}.
\vspace{-0.3cm}
\subsection{CCFM Evolution}
The CCFM\cite{ccfm} evolution equation is based on angular ordering
and colour coherence. As a result in the appropriate limit it reproduces
the DGLAP and the BFKL approximation. The angular ordering
of the emitted partons in the initial cascade
results from colour coherence and 
acts as unifying principle\footnote{From the proton $p$ to the photon
in the collinear DGLAP approximation $k_T$
and therefore the angle increases, while in the BFKL approximation
the angle increases because of the decreasing longitudinal momenta
($\theta \approx k_T/(z \, p)$). See ref.\cite{kimber} for an
introduction to angular ordering as interpolation between
DGLAP and BFKL.}.
At small values of the parton momentum fractions $z$, a random
$k_T$ walk is obtained. A first attempt\cite{smallx} to incorporate
the CCFM equations in a Monte Carlo simulation program 
was not suited for practical purposes.
Later a further analysis\cite{ccfmpheno} to make use of the CCFM equations
for small-$x$ phenomenology showed the importance of the inclusion
of soft corrections and revealed many theoretical 
difficulties, in particular with terms diverging for
high rather than low momentum fraction $z$.
Recently, significant progress has been made to correctly 
implement the CCFM evolution equations
and to efficiently generate unweighted events using the
backward evolution in Monte Carlo simulation 
program CASCADE\cite{cascade,ccfmhadfin}. 
A slightly different approach is followed in the LDC program\cite{ldc}. 
Both programs are able to make predictions for any physical observable
based on the hadronic final state. Only gluon-initiated processes
can be simulated up to now.
Some problems still remain, e.g. so far as in the CCFM evolution
only the singular terms in the splitting functions (for $z \to 0$ and $z \to 1$) 
have been considered.

\section{Small-$x$ Signatures in High Energy Collisions}
\label{sec:lowx}
Several signatures for small-$x$ QCD effects have been recently
investigated: jets
at large rapidity in $p\bar{p}$ collisions at
TEVATRON, the total hadronic cross section in $\gamma^* \gamma^*$
collisions at LEP and the rise of $F_2$ towards small-$x$
%the proton structure function
and the production of forward jets and particles in $ep$ collisions
at HERA. 

\subsection{Jet Production at Large Rapidity in $p\bar{p}$ Collisions}
\label{subsec:ppjet}
In the high energy limit of $p\bar{p}$ collisions 
%with a centre-of-mass (cm) energy $s$ 
inclusive dijet production (see Fig.~\ref{fig:lowxsig}b)
is sensitive to large logarithms of the form:
\begin{eqnarray}
\vspace{-0.3cm}
\sigma_{} \propto 
\exp{(\lambda \ln{\frac{W^2}{E_{T,1} \, E_{T,2}}})}
\propto \exp{(\lambda \, \Delta \eta )} \;
\label{eq:secpp} 
%\nonumber
\end{eqnarray}
where $Q^2 = E_{T,1}  E_{T,2}$ is the momentum transfer to the 
hard scattering and $W^2 = x_1 x_2 \, s$ 
is the squared cm energy of the hard scattering process\footnote{
Here $E_{T,i}$ are the transverse jet energies, $\eta_{i}$
the pseudo-rapidities 
and $x_{i}$ the momentum fractions of the two incoming partons.}.
The last proportionality in equation \ref{eq:secpp} follows after some
calculations for $\Delta \eta \gg 1$ from
\begin{eqnarray}
x_{i} = \frac{2 E_{T,i}}{\sqrt{s}} \, \exp{(\bar{\eta})} \, 
\cosh{\frac{\Delta \eta}{2}}
\label{eq:x2jet}
\end{eqnarray}
with $i=1,2$ and $\bar{\eta}= (\eta_1 + \eta_1)/2$ and $\Delta{\eta}= | \eta_1 - \eta_2 |$ for
the most backward ($\eta_1$) and the most forward ($\eta_2$) jet.
Equation \ref{eq:x2jet} is only valid for 2-body kinematics.

\begin{figure}
\vspace{-0.4cm}
\epsfxsize180pt
\figurebox{}{}{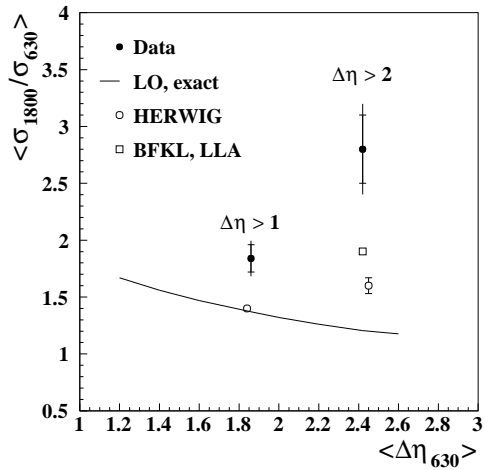}
\vspace{-0.2cm}
\caption{Ratio of the D0 dijet cross section at a cm
energy $\sqrt{s}=1800$ {\rm GeV} and $\sqrt{s}=630$ {\rm GeV}
for $\Delta \eta > 1$ and $\Delta \eta > 2$
as a function of the mean rapidity difference of the jets
at $\sqrt{s}=630$ {\rm GeV}.
}
\label{fig:d0_dsigdy}
\vspace{-0.4cm}
\end{figure}

The largest logarithms are therefore enhanced for
minimal jet $E_T$ and maximal $W$.
To avoid the strong dependence on the steeply falling
parton densities at small-$x$, it has been proposed 
by Mueller and Navelet\cite{mueller87} to study the dijet cross section 
at fixed $x_{i}$ at different cm energies $s$.
In an analysis performed by D0\cite{d0bfkl}, jets have been selected by
$E_{T,i} > 20$ {\rm GeV} and $| \eta_{i} | < 3$.
Moreover, $400 < ( Q^2 =   E_{T,1} E_{T,2} ) < 1000 ~{\rm GeV}^2$
was demanded. 
The cross section ratio at fixed $x_{i}$
for two cm energies has been measured:
\begin{eqnarray}
%R = \frac{\sigma_{}(\sqrt{s_a}=1.8 \, {\rm TeV})}
%         {\sigma_{}(\sqrt{s_b}=0.63\, {\rm TeV})}
%
R = \frac{\sigma_{}(\sqrt{s_a})}
         {\sigma_{}(\sqrt{s_b})}
  = \frac{\exp{(\lambda \, (\Delta \eta_a - \Delta \eta_b))}}
         {\sqrt{\Delta \eta_a/\Delta \eta_b}}
\label{eq:dijetratio} 
% \nonumber
\end{eqnarray}
The result of the cross section ratio at $\sqrt{s_a} = 1.8~{\rm TeV}$ 
and $\sqrt{s_b} = 630~{\rm GeV}$
is shown as a function of the mean $\av{\Delta \eta}$ for 
$\sqrt{s_b} = 630~{\rm GeV}$
in Fig.~\ref{fig:d0_dsigdy}.
At large $\Delta \eta>$ 
the dijet cross section increases almost by a factor of $3$
between the two cm energies. The strong increase is reflected
in a large mean $\lambda$ value: $\lambda = 0.65 \pm 0.07$,
stronger than expected by theoretical predictions\cite{orr98}.
The exact LO pQCD calculation leads to a falling cross section.
The LO BFKL calculation (labelled LLA in Fig.~\ref{fig:d0_dsigdy})
predicts $\lambda = 0.45$ 
for $\alpha_s(20 ~{\rm GeV})=0.17$. A complete NLO BFKL calculation
is not yet available. Surprisingly, the highest
$\lambda = 0.6$ is obtained by HERWIG, a LO QCD event generator
supplemented by leading log parton showers.

However, 
it has recently been pointed out\cite{delduca01} that an
interpretation of these results is difficult because of
differences in the definition of the cross sections between
the D0 data
and the original Mueller-Navelet proposal. This concerns the
reconstruction of the momentum fractions $x_{i}$ using eq.~\ref{eq:x2jet}
which is only valid for 2-body kinematics and the presence
of an upper bound on the momentum transfer $Q^2$.
These differences to the original proposal can be neglected
only in the asymptotic limit at large $s$
and large $\Delta \eta$. 
At TEVATRON, however, this limit is not reached.
Moreover, 
the requirement of two jets with the same minimum $E_T$
is particularly critical\cite{symcuts}, since
large logarithms not connected with BFKL effects make
an interpretation very difficult.
%%%%%%%%%%%%%%%%%%%%%%%%%%%%%%%%%%%%%%%%%%%%%%%%%%
%%%%%%%%%%%%%%%%%%%%%%%%%%%%%%%%%%%%%%%%%%%%%%%%%%%%
%
\subsection{Inclusive Hadronic Cross Section in 
$\gamma^* \gamma^*$ Collisions}
\label{subsec:gamgam}
By studying the total hadronic cross section 
$\gamma^* \gamma^* \to {\it hadrons}$ in $e^+ e^-$ collisions,
difficulties connected with the hadronic nature of incoming particles
can be naturally avoided. Virtual photons are colourless objects 
and their virtuality $Q^2$ controls the transverse
size $\propto 1/\sqrt{Q^2}$ of the hard processes.
For $Q^2 \gg \Lambda^2_{\rm QCD}$ 
a complete perturbative calculation is possible.
For small virtualities $Q_{i}^2$ of one of the virtual photons
and for large cm energies $W$ of the
$\gamma^* \gamma^*$ system the cross section contains large logarithms of the form:
%%%%%%%%%%%%%%%%%%%%%%%%%%%%%%%%%%%%%%%%%
\begin{eqnarray}
\sigma_{\gamma^* \gamma^*} \propto \exp{(\lambda \ln{\frac{W^2}{\sqrt{Q^2_1 Q^2_2}}})} 
= \exp{(\lambda \; Y}).
\nonumber
\label{eq:secgg} 
\end{eqnarray}
%%%%%%%%%%%%%%%%%%%%%%%%%%%%%%%%%%%%%%%%%
If $Q^2_1 \approx Q^2_2$, a DGLAP parton evolution
between the two photons is suppressed and a resummation
of the terms $Y = \ln{(W^2/\sqrt{Q^2_1 Q^2_2})}$
% = \ln{\frac{W^2}{\sqrt{Q^2_1 Q^2_2}}}$. 
is needed. 
%%%%%%%%%%%%%%%%%%%%%%%%%%%%%%%%%%%%%%%%%%%%%%%%%%
\begin{figure}[htp]
\vspace{-0.6cm}
\epsfxsize180pt
\figurebox{}{}{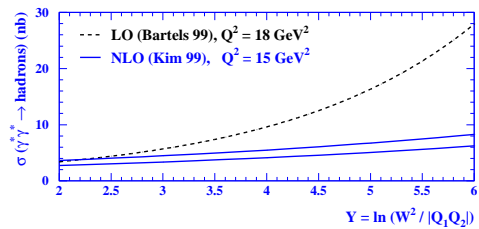}
%\vspace{0.3cm}
\vspace{-0.3cm}
\caption{Total hadronic cross section for the reaction
$\gamma^* \gamma^* \to X$ as a function of $Y$
calculated in a LO and an approximate NLO
BFKL calculation.
}
\label{fig:gammagammatheo}
\vspace{-0.4cm}
\end{figure}
%%%%%%%%%%%%%%%%%%%%%%%%%%%%%%%%%%%%%%%%%%%%%%%%%%%%
$\sigma_{\gamma^* \gamma^*}$ is therefore often
considered as a golden BFKL signature. 
An example of a Feynman diagram
is shown in Fig.~\ref{fig:lowxsig}c. 
As can be seen in Fig.\ref{fig:gammagammatheo},
in LO a strong increase of  
$\sigma_{\gamma^* \gamma^*}$ at high $Y$ is 
expected\cite{bartelgamgam}. An
approximation to a NLO calculation\cite{kim99} 
leads to a much suppressed cross section at high $Y$.
%
%BFKL calculation LO 
%includes effects of charm mass, running $\alpha_s$ and
%contribution of longitudinal photon polarisation states,
%is significantly larger than the data
% 
%%%%%%%%%%%%%%%%%%%%%%%%%%%%%%%%%%%%%%%%%%%%%%%%%%
\begin{figure}
\vspace{-0.3cm}
\epsfxsize190pt
\figurebox{}{}{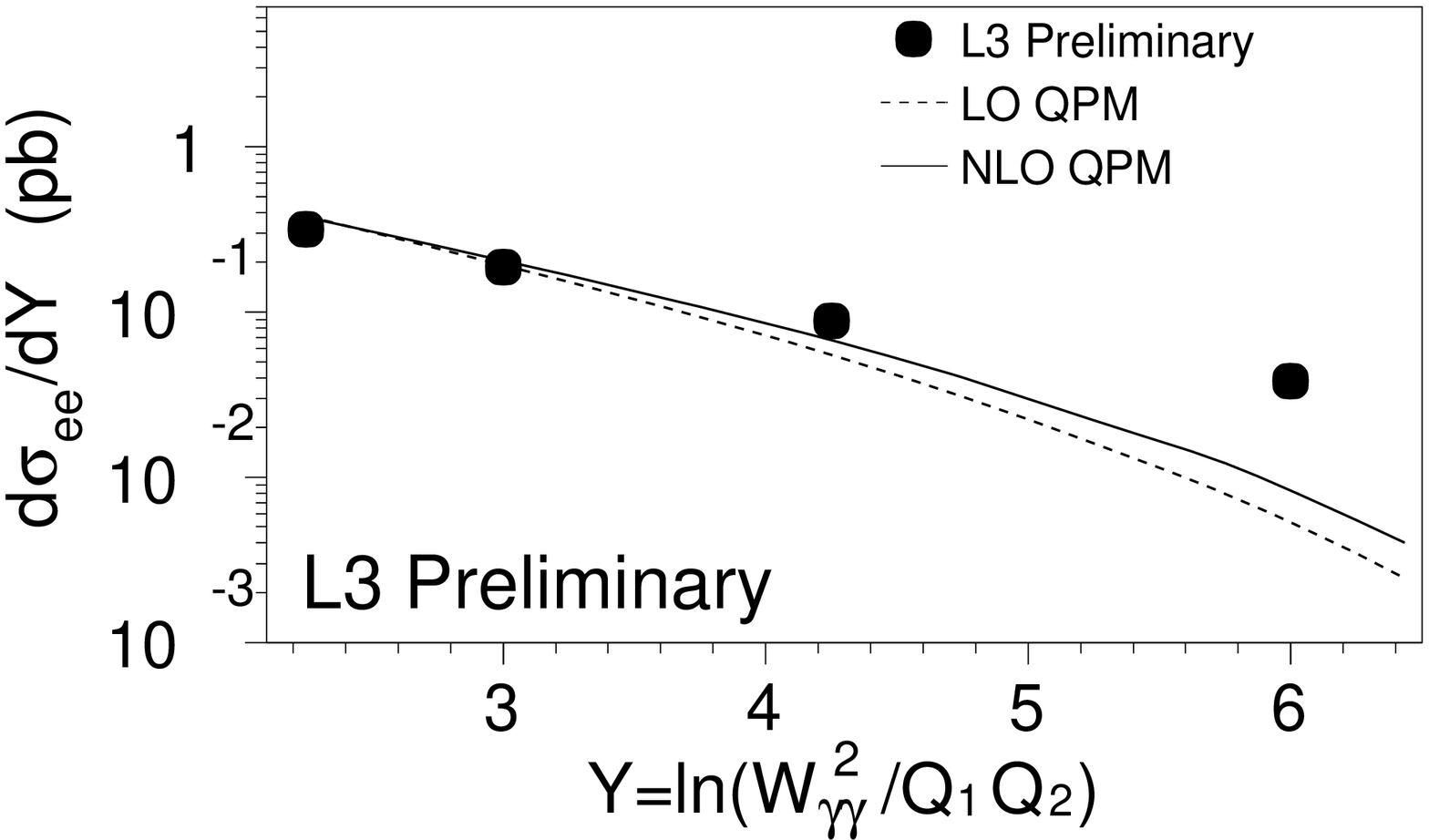}
\epsfxsize200pt
\figurebox{}{}{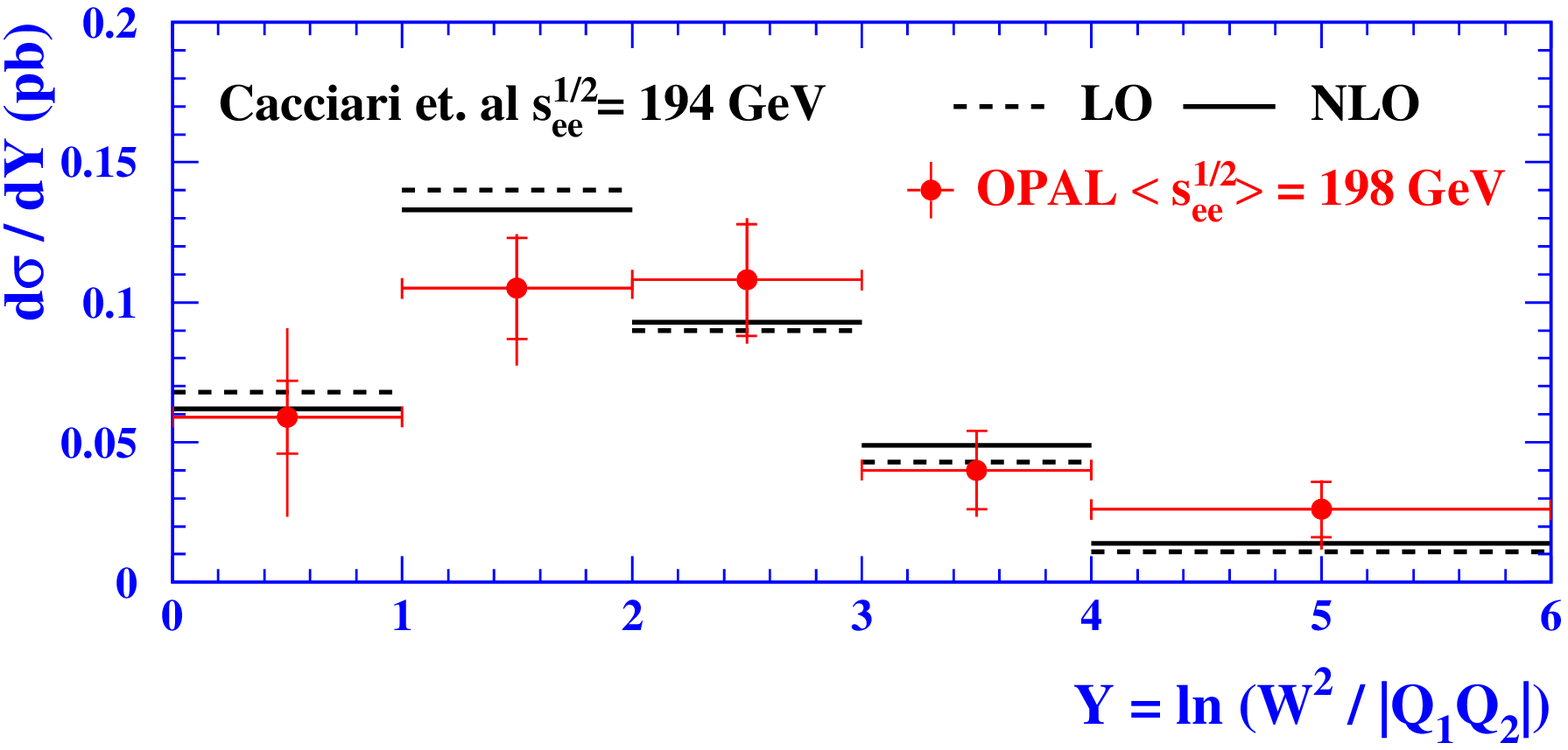}
\begin{picture}(50.,50.)
\put(5.,160) {a)}
\put(5.,50) {b)}
\end{picture}
\vspace{-1.7cm}
\caption{Total hadronic cross section 
$\gamma^* \gamma^* \to X$ as a function of 
the variable $Y$ a) measured by L3 
and b) by OPAL. Overlayed is a LO and NLO fixed order 
calculation.
}
\label{fig:l3_dsigdy}
\vspace{-0.6cm}
\end{figure}
%%%%%%%%%%%%%%%%%%%%%%%%%%%%%%%%%%%%%%%%%%%%%%%%%%%%

In a first measurement of $\sigma_{\gamma^* \gamma^*}$ 
in $e^+e^-$ collisions at $\sqrt{s_{ee}}=91-183$ {\rm GeV}
the predicted significant rise towards large $Y$
was observed by L3\cite{l3}. 
%
%the data are significantly higher than the prediction without
%the BFKL resummation. However, the LO BFKL prediction significantly
%overshoots the data. The power of the growth of the cross section
%is determined to $\lambda= 0.29 \pm 0.025$ for scales in the range 
%$3.5 - 14 {\rm GeV}^2$. 
%
However these data have not been corrected
for QED radiative effects. In particular, if the
kinematics is reconstructed using the scattered 
electrons, these corrections are substantial.
In a recent analysis\cite{l3_01}, L3 has presented cross sections corrected
for radiative effects using $e^+ e^-$ data
at a cm energy of $\sqrt{s_{ee}} = 189 - 209$~{\rm GeV}. 
The QED corrections lower the measured uncorrected cross section
by $58\%$ at large $Y$ and, moreover, can be very different for
the various contributing QCD subprocesses.
The cross section measured in the kinematic region:
$4 < Q^2_{i} < 44~{\rm GeV}^{2}$, $W > 5$~{\rm GeV},
electron energy $E_{i} > 40$~{\rm GeV} and the polar electron 
angle $30 < \theta_{i} < 66$~mrad is shown in
Fig.~\ref{fig:l3_dsigdy}a). A fixed order calculation
in LO as well as in NLO\cite{cacciari01} is able to describe the
data at low $Y$ ($Y < 5$). 
However, at the largest $Y$ ($5 < Y < 7$), the calculations are below
the data by $4$ standard deviations.
The NLO corrections are relatively small and
they only slightly increase the cross section at large $Y$. 
L3 concludes
that ``this can indicate the presence of QCD resolved processes
or the onset of BFKL dynamics''\cite{l3_01}. 

However, this effect has not been confirmed by OPAL\cite{opal}
analysing $e^+e^-$ collisions at cm energies of $\sqrt{s_{ee}}=189-209$ {\rm GeV}.
The measured cross section $\sigma_{\gamma^* \gamma^*}$
(for $E_{i} > 0.4 \, E_b$, where $E_b$ is the beam energy, 
$34 < \theta_{i} < 55$~mrad
and $W > 5$~{\rm GeV}) for an average $Q^2 = 17.9~{\rm GeV}^2$ 
is shown in Fig.~\ref{fig:l3_dsigdy}b) as a function of $Y$.
The exact LO and the NLO calculation reproduce the data well. Also an
approximate higher order BFKL calculation\cite{motyka}  
imposing energy momentum conservation along 
the gluon ladder, is in agreement with the data (not shown).
%
%Also the event generator
%PHOJET\cite{phojet} based on LO QCD for the hard processes and
%a generalised vector dominance model for soft processes
%describes the data.
%data
%
Although in the L3 analysis slightly higher $W$ and $Y$ values
are reached, the different conclusions which can be drawn
from the two analyses is surprising. Some
experimental questions in particular the control of the
radiative corrections as a potentially large source of
systematic uncertainty needs to be 
rediscussed\footnote{One question is e.g. to which extent the
size of the QED corrections depend on the 
Born cross section itself. An iterative procedure 
might be required.} % to extract $\sigma_{\gamma^* \gamma^*}$.} 
before firm conclusions can be drawn.
%However, in both analyses the strong increase expected by the
%LO BFKL calculation has not been observed. 
To unambiguously establish small-$x$ effects in $\gamma^*  \gamma^*$
collisions  
one probably needs much higher cm energies, like the one that will be
available at TESLA.

\subsection{Inclusive DIS: $F_2$}
\label{subsec:f2}
A classic key measurement to understand the structure of the proton
and its dynamical processes 
is the precise determination of the structure function
$F_2$ by counting inclusively the lepton scattered off the proton.
The HERA measurements\cite{F2,F2new} span about $6$ orders of magnitude
in $Q^2$, $ 0.1 \lapprox Q^2 \lapprox 30000 ~{\rm GeV}^2$,
and cover the momentum range from the sea quark region
at low $x$ ($ x \gapprox 10^{-6}$) 
to the valence quark region at large $x$ ($x \lsim 0.65$).

%%%%%%%%%%%%%%%%%%%%%%%%%%%%%%%%%%%%%%%%%%%%%%%%%%
\begin{figure}%1
\epsfxsize200pt
\figurebox{}{}{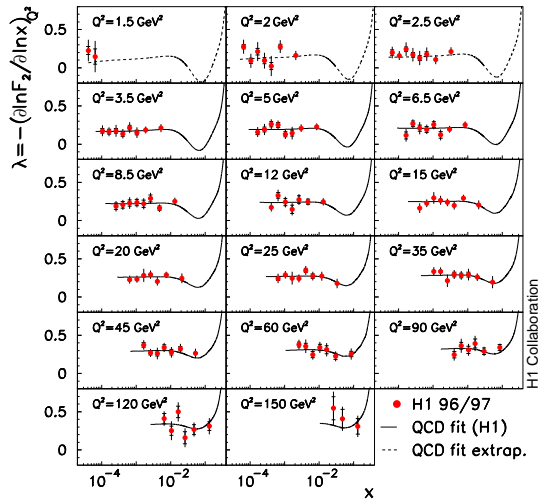}
\vspace{-0.45cm}
\caption{Slope of the %inclusive total $gamma$p cross section 
proton structure function $\lambda$ as a function of $x$ in bins of $Q^2$.
}
\label{fig:h1f2slope}
\vspace{-0.4cm}
\end{figure}
%%%%%%%%%%%%%%%%%%%%%%%%%%%%%%%%%%%%%%%%%%%%%%%%%%%%

Compared to observables based on the hadronic final state, 
the measurement of $F_2$ has the advantage that it
can be directly compared to pQCD calculations.
However, because of its inclusive nature, 
small novel effects on top of the
dominating DGLAP parton evolution might be difficult to reveal.
Therefore $F_2$ has to be precisely pinned down to get
a handle on the QCD evolution and to simulanteously constrain
the non-perturbative parton density functions.
In the most recent $F_2$ measurements a precision of $2-4\%$
in the high statistics region $Q^2 \lsim 100~{\rm GeV}^2$ 
has been achieved\cite{F2new}.
One of the most important results is that
the rise of $F_2$ towards small-$x$: 
%%%%%%%%%%%%%%%%%%%%%%%%%%%%%%%%%%%%%%%%%%
\vspace{-0.3cm} 
\begin{eqnarray}
\sigma_{ep} \propto F_2 \propto
\exp{(\lambda \ln{\frac{W^2}{Q^2}})} \approx x^{-\lambda}
% \nonumber
\label{eq:secdis} 
\vspace{-0.6cm} 
\end{eqnarray}
%%%%%%%%%%%%%%%%%%%%%%%%%%%%%%%%%%%%%%%%%%
expected in the DLL or BFKL limit has been observed. 
However, analyses based on the NLO DGLAP equations give a
satisfactory description of the data down to remarkably
low $Q^2$ and $x$ values\cite{DGLAPfit,mrst,cteq}.
In these analyses the quark and gluon densities are parameterised
at a starting scale and are adjusted such that the data are described.
While $F_2$ constrains mainly the quark densities, the scaling violation
$d F_2/d \log{Q^2}$ give access to the gluon density. 
Whether this proves the correctness of the DGLAP picture
or is just due to a large flexibility of the fit, in particular
in the parameterisation of the parton density function,
can not yet be answered.

\begin{figure}
\vspace{-1.cm}
\epsfxsize175pt
\figurebox{120}{120}{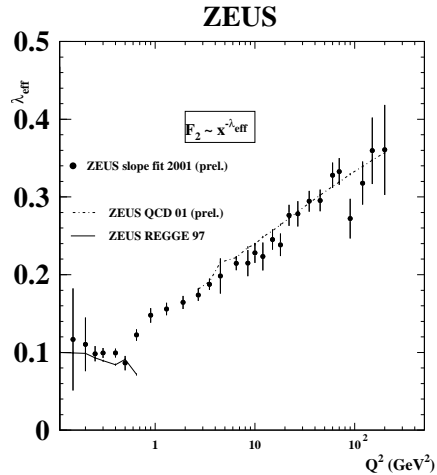}
\vspace{-0.1cm}
\caption{Slope of the incl. total $\gamma^{(*)}$p cross
section~$\lambda$ as a function of $Q^2$. 
Shown is a generalised vector dominance model
and a QCD NLO DGLAP fit. 
}
\label{fig:lambda_zeus}
\vspace{-0.6cm}
\end{figure}

To improve the sensitivity to small-$x$ QCD effects, 
the H1 collaboration\cite{h1f2slope}
has measured the slope $dF_2/d\ln{x}$ for fixed $Q^2$ 
as a function of $x$ (see Fig.~\ref{fig:h1f2slope}).
%In this way the strong $Q^2$ dependence of $\lambda$ 
%($\lambda \approx 0.3-0.4$ at $Q^2 \approx 100~{\rm GeV}^2$
%and $\lambda \approx 0.15$ at $Q^2 \approx 1.5~{\rm GeV}^2$.
%is removed. 
%For fixed $Q^2$,
The slope does not change with $x$ and the NLO DGLAP fit describes the data.
$dF_2/d\ln{x}$ can therefore be identified with the slope $\lambda$ only
changing with $Q^2$.
The $Q^2$ dependence of $\lambda$ is even more visible 
in the ZEUS measurement shown
in Fig.~\ref{fig:lambda_zeus} (see also sec. \ref{sec:diffraction}).
%
%%%%%%%%%%%%%%%%%%%%%%%%%%%%%%%%%%%%%%%%%%%%%%%%%%
%%%%%%%%%%%%%%%%%%%%%%%%%%%%%%%%%%%%%%%%%%%%%%%%%%%%
%
Despite this success it is worth noting that
in most NLO DGLAP fits %\cite{DGLAPfit,F2new,mrst,cteq}
%the extracted gluon density tends to go valence-like or negative at  
%$Q^2 \sim 1 ~{\rm GeV}^2$ indicating that the applicability of the
%DGLAP approach becomes questionable. Even more serious is that also 
the longitudinal structure function $F_L$ becomes
negative, which is obviously unphysical.
The inclusion of $\ln{(1/x)}$ terms 
is able to cure these problems\cite{f2bfkl,thorne,alta01}.
%
%%%%%%%%%%%%%%%%%%%%%%%%%%%%%%%%%%%%%%%%%%%%%%%%%%
\begin{figure}
\vspace{-0.4cm}
\epsfxsize200pt
\figurebox{}{}{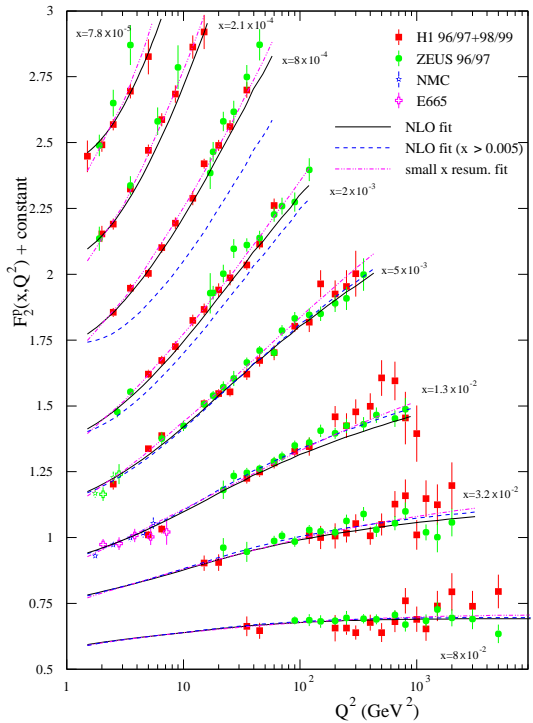}
\vspace{-0.4cm}
\caption{Proton structure function data as a function of
$Q^2$ at fixed $x$ compared with several different fits
within the MRST global fit framework.}
\label{fig:thorne}
\vspace{-0.7cm}
\end{figure}
%%%%%%%%%%%%%%%%%%%%%%%%%%%%%%%%%%%%%%%%%%%%%%%%%%%%
% MRST is global fit where
% fixed target experiment: 
% {u},{d}, \bar{u}, \bar{d}, {s}, \bar{s} at x \sim 0.1 $Q^2 \approx 10$
% HERA:
% sea quarks and dF2/dlogQ2 determines gluon x \sim 10^-3 $Q^2 \approx 10$
% TEVATRON: quark/gluon for x\sim 0.1 and $Q^2 \approx 10^4$ GeV2
%
%
If one aims for a global description of all available
data, the standard NLO DGLAP fit has to be stretched to the edge\cite{mrst01}.
In particular, there is a complicated
interplay between the small-$x$ DIS data and the high-$x$ 
inclusive jet cross sections from TEVATRON\cite{d0jet,cdfjet}.
The jet data prefer a high gluon at large $x$ values ($x > 0.1$)
and this significantly influences
the small-$x$ gluon via the momentum sum rule.
%
%With the high-$x$ gluon constraint the DIS data prefer a rather high
%value of $\alpha_s = 0.121$. However, the jet data alone prefer 
%$\alpha_s = 0.115$. Forcing $\alpha_s$ to be lower deteriorates
%the fit to the DIS data, since then they prefer a smaller gluon at high $x$ values.
%
The DIS data prefer a smaller gluon at high-$x$ values.
Therefore, some kind of compromise has to be found. Within the present
accuracy of the data, this is still possible, but is seems
that there is not much freedom left in the NLO DGLAP fit.

Within the framework of the MRST global QCD analysis\cite{mrst,mrst01}, 
it has been shown that problems
in describing HERA $F_2$ data at medium-$x$ values ($\approx 10^{-2}$) 
can be avoided, if the data at small-$x$ ($x \lsim 10^{-3}$)  
are not considered in the fit or, alternatively, if a small-$x$ resummation
obtained by solving the running coupling BFKL equation
is included\cite{thorne,thorne2}. 
This is demonstrated in 
Fig.~\ref{fig:thorne}\footnote{I thank R. Thorne for providing me with this figure.} 
where the standard NLO DGLAP fit gives a bad description of the data at
$x = 1.3 \cdot 10^{-2}$ and $100 \lsim Q^2 \lsim 1000~{\rm GeV}^2$ 
while a fit only including data with $x \gsim 5 \cdot 10^{-3}$
is much better. This fit, however, does not describe the small-$x$ data.
When small-$x$ resummation is included the data 
can be described over the whole kinematic domain.

In conclusion, the NLO DGLAP fit describes DIS data
to surprisingly small-$x$ and $Q^2$ values. However, there 
are some indications that the flexibility in the input parameters
is stretched to the edge, in particular, if the new inclusive high $E_T$
jet data from TEVATRON are included.
Fits including small-$x$ resummation lead to an improved description
of the data and a better convergence of $F_L$ when going from LO
to NLO.
\vspace{-0.6cm}
\subsection{Heavy Quark Production }
\label{sec:heavy}

The production of heavy quarks $Q$ in DIS
is directly sensitive to the
gluon content of the proton, since in LO the 
process $\gamma g \to Q \bar{Q}$ dominates. 
%(see Fig.\ref{fig:bbgraph}). 
Heavy quark cross sections are therefore an excellent testing
ground for the dynamics of the parton evolution.
%
%%%%%%%%%%%%%%%%%%%%%%%%%%%%%%%%%%%%%%%%%%%%%%%%%%
%\begin{figure}
%\epsfxsize200pt
%\vspace{-2.cm}
%\figurebox{}{120pt}{}
%\vspace{-1.8cm}
%\caption{Feynman graph for the production of heavy
%quarks in $ep$ collisions.
%}
%\label{fig:bbgraph}
%\vspace{-0.3cm}
%\end{figure}
%%%%%%%%%%%%%%%%%%%%%%%%%%%%%%%%%%%%%%%%%%%%%%%%%%
The inclusive $D^*$ meson cross section has recently been
measured by H1\cite{h1charm}
in the range $1 < Q^2 < 100~{\rm GeV}^2$, $0.05 < y < 0.7$,
$p_t^{D^*} > 1.5$ {\rm GeV} and $|\eta^{D^*}| < 1.5$ to be:
%\begin{equation}
$
\sigma_{vis}  = 8.5 \pm 0.42 ({\rm stat})
 ^{+1.02}_{-0.76} ({\rm syst}) \pm 0.65 {\rm (model)}~{\rm nb}.
%\end{equation}
$
A NLO calculation\cite{hvqdis} based on the heavy quark
matrix elements, DGLAP parton evolution and the
Peterson fragmentation function\cite{peterson}
only predicts $\sigma_{vis}= 5.1 - 7$ nb depending
on the exact value of the charm quark mass $m_c$ and the
Peterson fragmentation parameter $\epsilon$.
The most significant difference between data and NLO
is observed towards the proton direction, i.e. at large $\eta$.
Similar results have been obtained by ZEUS\cite{zeuscharm}. 
A possible explanation that this effect is due to 
neglecting the colour force between the charm quark 
and the proton remnant\cite{norbin}
is not large enough to explain this discrepancy.
The CCFM prediction, based on the off-shell matrix elements
and on an unintegrated gluon density obtained from a $F_2$ fit, 
has been calculated using CASCADE\cite{cascade} and 
agrees with the data.

The visible cross section can be converted to the semi-inclusive
charm structure function $F_2^c$ which is easier to compare
to theory predictions. However, the extraction of  $F_2^c$
faces the problem that only part of the total cross
section (typically $30\%$)
can be measured in the detector. 
%%%%%%%%%%%%%%%%%%%%%%%%%%%%%%%%%%%%%%%%%%%%%%%%%%
\begin{figure}
\epsfxsize200pt
\vspace{-1.cm}
\figurebox{120}{120}{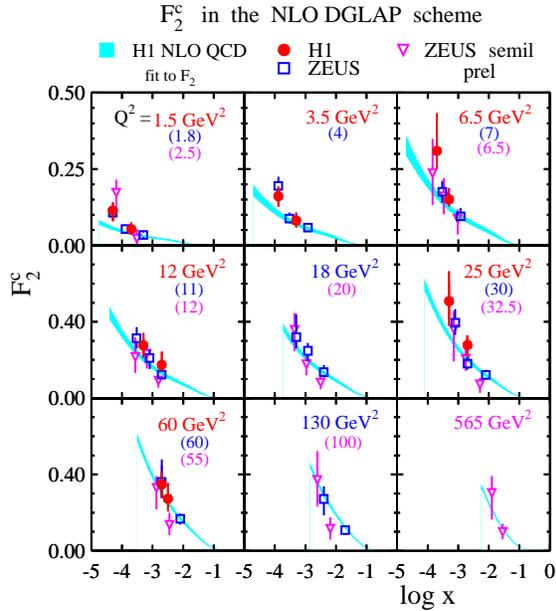}
\vspace{-1.8cm}
\caption{Charm proton structure function $F_2^c$ as
a function of $\log{x}$ in $Q^2$ bins.
}
\label{fig:f2c}
\vspace{-0.5cm}
\end{figure}
%%%%%%%%%%%%%%%%%%%%%%%%%%%%%%%%%%%%%%%%%%%%%%%%%%
This quantity is therefore significantly biased by the 
theoretical model used to determine the experimental acceptance 
(up to $20\%$ at the smallest $x$).
$F_2^c$ extracted in the NLO DGLAP scheme is shown in 
Fig.~\ref{fig:f2c}. A steep rise towards small-$x$ 
and a clear dependence of the slope $\lambda$ on $Q^2$
is observed. In particular, in the low $Q^2$ region
this rise tends to be steeper than predicted by the
NLO DGLAP calculation\footnote{This is true when compared to
the H1 NLO DGLAP fit. In the ZEUS fit a slightly larger $F_2^c$ error band
is obtained making the above conclusion less clear.}.
At small-$x$, $F_2^c$ grows at the same rate as
the fully inclusive structure function $F_2$ as is
expected, since both processes are initiated by gluons.

%%%%%%%%%%%%%%%%%%%%%%%%%%%%%%%%%%%%%%%%%%%%%%%%%%
\begin{figure}
\epsfxsize220pt
\vspace{-0.5cm}
\figurebox{120}{120}{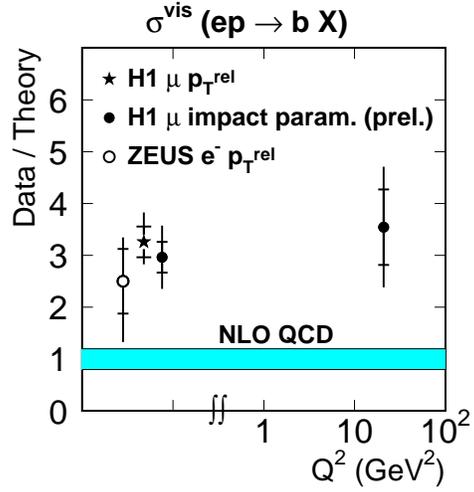}
\vspace{-0.9cm}
\caption{Ratio of the measured to the calculated
total beauty production cross section in
$e p$ collisions. 
}
\label{fig:epbeauty}
\vspace{-0.3cm}
\end{figure}
%%%%%%%%%%%%%%%%%%%%%%%%%%%%%%%%%%%%%%%%%%%%%%%%%%%%

This year H1 has reported\cite{h1beautydis} on the 
first measurement of beauty production
in DIS in the region $2 < Q^2 < 100~{\rm GeV}^2$, $0.05 < y < 0.7$ for muon
polar angles $35^o < \theta < 130^o$ and transverse momenta 
$p_T > 2$ {\rm GeV}. The beauty cross section is derived by determining the
fraction of beauty events decaying semi-leptonically to muons
using the transverse momentum of the muon with respect to the jet axis ($P_{T,rel}$)
and by exploiting the finite lifetime of b-mesons by measuring
the distance of closest approach of the muon track with respect
to the primary event vertex (impact parameter).
The measured cross section 
$\sigma_{ep \to b X \to \mu X} = 39 \pm 8 ({\rm stat}) \pm 10 ({\rm syst}) {\rm pb}$ 
is significantly above the NLO prediction\cite{hvqdis} of
$\sigma~=~11~\pm~2 {\rm pb}$.
In the calculation, a beauty mass $m_b = 4.75$~{\rm GeV}, 
a renormalisation and factorisation scale 
$\mu = \sqrt{Q^2+ 4 m_b^2}$ and a Peterson fragmentation parameter
$\epsilon=0.0033$ are used. The LO cross section is only $\sigma= 9$~{\rm pb}.
The CASCADE program based on the CCFM equation predicts $\sigma = 15$~{\rm pb}.
A similar excess has been reported in quasi-real photon-proton collisions by H1\cite{h1beauty},
where the cross section for $b \bar{b}$ production 
for $Q^2 < 1~{\rm GeV}^2$ is measured.
A cross section value larger than the theory prediction is also reported
by the ZEUS collaboration\cite{zeusbeauty} in the semi-leptonic electron channel.
The $b \bar{b}$ cross section for $Q^2 < 1~{\rm GeV}^2$,
$0.2 < y < 0.8$, $p_{t,b} > 5$~{\rm GeV} and $|\eta_b| < 2$ is:
$\sigma_{ep \to e b \bar{b} X} = 1.6^{+0.54}_{-0.75} {\rm nb}$.
The NLO prediction is only $0.64$~{\rm nb}.

%%%%%%%%%%%%%%%%%%%%%%%%%%%%%%%%%%%%%%%%%%%%%%%%%%
\begin{figure}
\epsfxsize160pt
\vspace{-0.8cm}
\figurebox{120}{120}{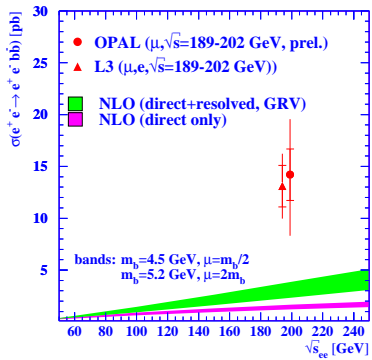}
\vspace{-0.3cm}
\caption{Total beauty production cross section in
$\gamma \gamma$ collisions as a function of the
cm energy $\sqrt{s_{ee}}$.
}
\label{fig:ggbeauty}
\vspace{-0.8cm}
\end{figure}
%%%%%%%%%%%%%%%%%%%%%%%%%%%%%%%%%%%%%%%%%%%%%%%%%%
A comparison between data and theory for the various measurements is shown
in Fig.~\ref{fig:epbeauty}. For all measurements a higher cross section 
with respect to the NLO predictions is measured.
However, the total cross section measurement is affected by large extrapolation factors
calculated using LO MC simulation programs. For instance, 
jets with $E_T > 6$~{\rm GeV} have to be selected to calculate $P_{T,rel}$.
In the quoted total cross section they are not included. 
Only $42.9\%$ of the cross section is really
measured in the detector\cite{h1beauty}. 
Moreover, requiring two jets with equal $E_T$ is
problematic\cite{symcuts}. The extrapolation includes
large model uncertainties\cite{zeusbeauty,jungring} (up to $30\%$). 
It is interesting that the
ZEUS measurement, where also the beauty cross section including the jet requirement
is quoted, agrees well with the CASCADE result and with LO QCD MC simulations, if heavy
quark excitation processes possibly mimicking part of the 
NNLO corrections\cite{norbin} are included.
 
The increase of the cm energy of LEP to 
$\sqrt{s_{ee}} \approx 200$~{\rm GeV} 
has also made possible the first
observation of beauty production in $\gamma^* \gamma^*$ collisions. Here also an
excess of the measured beauty cross section over the theory prediction
is found\cite{beautylep} (see Fig.~\ref{fig:ggbeauty}).
The NLO prediction agrees with the measured charm cross
sections, but in the case of beauty it underestimates the cross section by about two
standard deviations.
Since many of the phenomenological challenges are similar to $ep$ collisions,
e.g. the description of heavy quarks near the production threshold or
the treatment of the hadronic structure of the photon, this indicates
that the excess of beauty events at HERA is not accidental, but 
might have a deeper reason.

%%%%%%%%%%%%%%%%%%%%%%%%%%%%%%%%%%%%%%%%%%%%%%%%%%
\begin{figure}
\epsfxsize180pt
%\vspace{-2.cm}
\figurebox{120}{120}{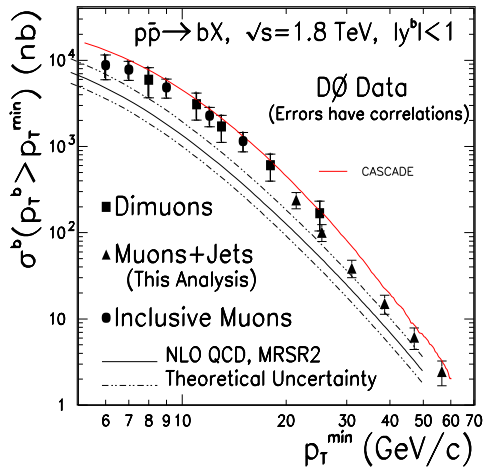}
\vspace{-0.2cm}
\caption{
Beauty production cross section for different 
minimal values of the transverse beauty momentum
in $p\bar{p}$ collisions at $\sqrt{s}=1800$ {\rm GeV}. 
}
\label{fig:d0_bbar}
\vspace{-0.3cm}
\end{figure}
%%%%%%%%%%%%%%%%%%%%%%%%%%%%%%%%%%%%%%%%%%%%%%%%%%%%
Also in $p\bar{p}$ collision a higher beauty cross section has been measured\cite{tevatronbeauty} 
than expected by NLO calculations\cite{tevatronnlo}. 
In general, the observed excess grows towards
large rapidities, i.e. towards the proton remnants.
One of the measurements performed in the central rapidity region
is shown in Fig.~\ref{fig:d0_bbar} over a wide minimum transverse momentum range $p_{T,b}^{\rm min}$.
Shown is the b-meson cross section (incl. muon) as well as a jet cross section
where the jets contain a b-tag (muon+jet). 
For the last observable the treatment 
of soft and collinear divergencies is safer in the theoretical calculation\cite{frixione97}.
For both observables the data lie higher than the theory prediction.
Even so, the agreement with the theory is improved in the latter case
in particular at high $p_{T,b}^{\rm min}$, 
a discrepancy still remains with respect to the central result.

A better agreement with the data is found using CASCADE based on the CCFM
equations\cite{junghq} (see Fig.~\ref{fig:d0_bbar}). This approach
uses LO off-shell matrix elements for heavy quarks and the scale is set
to $\mu^2 = m_b^2+p_{T,b}^2$. In addition, the unintegrated gluon density,
which has been obtained from an input parameterisation using only
two parameters adjusted to HERA data, enters the calculation
\footnote{At the TEVATRON cm energy, the gluon-gluon fusion is the dominant production mechanism.}. 
Note, that in a typical NLO DGLAP fit much more free parameters are used
to parameterise the parton densities. 
%This unintegrated gluon distribution is flat at the starting scale. 
In view of the few input parameters only fixed by
DIS data the agreement of the CASCADE calculation 
with the data is quite remarkable.
A similar success in describing beauty production in 
$p\bar{p}$ collisions has been obtained using
$k_T$ factorisation, an unintegrated
parton distribution and a BFKL vertex for the $q \bar{q}$-vertex\cite{hagler}.

Whether the higher beauty cross sections seen in various processes really can be
attributed to the necessity to modify the parton evolution at small-$x$\footnote{
Note that the $x$ values in some of the observables are not really small.
Moreover, the work to include finite $z$ terms in CASCADE is still in progress.}
or whether it is just due to other complications like the resummation
of large logarithms in the presence of two or more hard scales or problems
connected to the modelling of the fragmentation
cannot be answered today.
The larger amount of data which will be available after
the HERA and TEVATRON luminosity upgrades together with the improved
efficiency of the upgraded detectors will allow 
further improvements in the experimental precision and to extend the measurement 
towards larger $p_T$ and larger $Q^2$.
This might allow possible solutions to this problem to be disentangled. 

\vspace{-0.3cm}
\subsection{Forward Jet and Particle Production in DIS}
\label{sec:fwdjet} 
\begin{figure}%1
\epsfxsize200pt
\figurebox{}{}{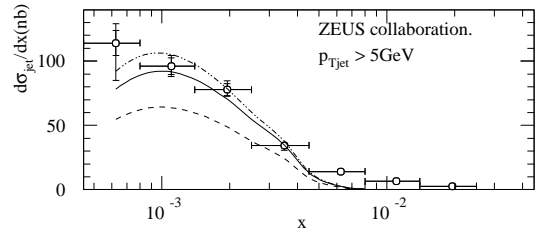}
\vspace{-0.5cm}
\caption{Forward jet DIS cross section as function of $x$. 
Approx. NLO BFKL calculations for different scales 
and infra-red cut-offs are shown as lines.
}
\label{fig:fj94}
\vspace{-0.6cm}
\end{figure}

The classical signature designed to enhance BFKL effects 
 in DIS is the production
 of ``forward jets''\cite{fjets} characterised by $k_T^2 \approx Q^2$
 and $x_{\rm jet}=E_{\rm jet}/E_p$ as large and $x$~as small
 as kinematically possible (see Fig.\ref{fig:lowxsig}d)\footnote{   
 $E_{\rm jet}$ ($E_p$) denotes the energy of the forward jet (proton)
 in the laboratory system.}.
 The first requirement suppresses the $k_T$ ordered DGLAP evolution.
 When asking for large $x_{\rm jet}/x$,
 the forward jet is separated by a large rapidity
 interval from the struck quark such that the phase space
 for parton emissions between the two is amplified.  
 In this case the $\alpha_s \ln{(x_{\rm jet}/x)}$ 
 terms are expected to become so large that their resummation 
 leads to a sizable increase of the forward jet cross-section: 
%%%%%%%%%%%%%%%%%%%%%%%%%%%%%%%%%%%%%%
\begin{eqnarray}
\vspace{-0.8cm}
\sigma_{\footnotesize BFKL} \propto 
\exp{(\lambda \ln{\frac{x_{jet}}{x}})}. 
 \nonumber
\label{eq:secfwdjet} 
\end{eqnarray}
%%%%%%%%%%%%%%%%%%%%%%%%%%%%%%%%%%%%%%
A fast rise of the cross-section for forward jets 
with decreasing $x$ has indeed been seen in the data\cite{fjetdata}.
A LO BFKL calculation\cite{fjetbfkl} rises steeper than
the data. A more recent calculation emulating some of the expected
NLO effects\cite{bfklfwdjet} is in much better agreement (see Fig.\ref{fig:fj94}).
Also CASCADE based on the CCFM equations reproduces
the data\cite{ccfmhadfin}.

%%%%%%%%%%%%%%%%%%%%%%%%%%%%%%%%%%%%%%%%%%%%%%%%%%
%%%%%%%%%%%%%%%%%%%%%%%%%%%%%%%%%%%%%%%%%%%%%%%%%%%%

%However, typical $z$ values at HERA, for instance, in the forward
%jet regime (see section \ref{subsec:intro}) are around $z \sim 0.5$ and higher.
%Therefore non-singular terms in the splitting functions, 
%which are missing in the present implementations of the CCFM and
%the BFKL equations, have to be considered \cite{jungdis00}.

The good description of the forward jet cross
section can be taken as an indication of the need
for the BFKL resummation in an extreme
phase space, where small-$x$ effects are expected
to be important. However, other interpretations are possible.
For instance, a calculation
including direct and resolved photon interaction also 
gives a good description of the data\cite{jungresolved,kramer99}. 
Kramer and P\"otter
argue that the
success of their NLO calculation can be taken as a strong
indication that a direct ${\cal O}(\alpha_s^3)$ calculation, i.e.
without the need to introduce a virtual photon structure,
would also describe the data.
In this case the need for a BFKL resummation is just not
strong enough at HERA to be unambiguously distinguished from an
exact fixed order calculation.
However, a recent H1 analysis\cite{schoerner} systematically investigating
the phase space regions from a simple inclusive jet
cross section to the forward jet regime shows that neither
a direct NLO calculation nor the NLO calculation
including resolved photons describes
the data in the whole phase space. While e.g.
the forward jet cross section is described, the
calculation overshoots the simple single inclusive
jet cross section in the backward region.

In conclusion,
despite the wealth of jet data at small-$x$ which became recently
available a consistent physical interpretation 
has not been achieved.

\begin{figure}
\vspace{-0.4cm}
\epsfxsize150pt
\figurebox{}{}{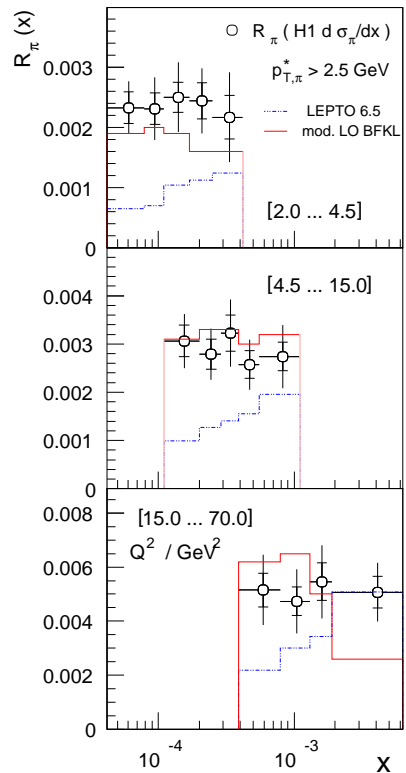}
%\vspace{0.3cm}
\caption{Rate of forward $\pi^0$ to the inclusive cross section as a function of $x$.
}
\label{fig:h1pi}
\vspace{-0.7cm}
\end{figure}

A complementary approach to gain insights in the parton evolution
is the production of particles at $E_T$\cite{kuhlen}. 
Charged or neutral particles at high $E_T$ are well correlated to 
parton activity. Their production rate can be directly
predicted by analytic calculations folded with proper fragmentation
functions. The measured $\pi^0$ cross section 
exhibits the same rise towards small-$x$ as the inclusive one,
i.e. the ratio $R_\pi(x)$ is constant
(see Fig.~\ref{fig:h1pi})\cite{fwdpizero}. 
At $2 < Q^2 < 4.5~{\rm GeV}^2$ approximately $0.2\%$
of the DIS events contain a forward $\pi^0$ with 
$E_T > 2.5$ {\rm GeV}. At $15 < Q^2 < 70 {\rm GeV}^2$, 
this ratio rises to $0.5 \%$.
The MC based on LO QCD and parton showers
(labelled LEPTO) cannot describe this behaviour, while a 
BFKL calculation approximating some of the NLO effects by requiring
energy-momentum conservation and allowing for a running
coupling describes the data\cite{bfklfwdjet}.

%%%%%%%%%%%%%%%%%%%%%%%%%%%%%%%%%%%%%%%%%%%%%%%%%%
%%%%%%%%%%%%%%%%%%%%%%%%%%%%%%%%%%%%%%%%%%%%%%%%%%%%

%%%%%%%%%%%%%%%%%%%%%%%%%%%%%%%%%%%%%%%%%%%%%%%%%%%%
%\begin{figure}%1
%\epsfxsize200pt
%\vspace{-2.cm}
%\figurebox{120}{120}{f2_x.ps}
%\vspace{-1.8cm}
%\caption{Proton structure function as a function of $x$ in
%bins of $Q^2$. Was bedeuten die Kurven?
%}
%\label{fig:f2x}
%\vspace{-0.3cm}
%\end{figure}
%%%%%%%%%%%%%%%%%%%%%%%%%%%%%%%%%%%%%%%%%%%%%%%%%%%%
\vspace{-0.5cm}
\section{Inclusive and Diffractitive DIS, 
Transition from Soft to Hard}
\label{sec:diffraction}
\vspace{-0.3cm}
The constant (with respect to the inclusive cross section) 
probability to emit hard particles in the forward region
can be juxtaposed to the probability to emit
no particle at all. In approximately $5-10 \%$ of all DIS events 
no particle is measured in a large rapidity region\cite{diffhera}.
The latter class of events are called ``diffractive''\footnote{
In the analogy of the diffraction of light, see\cite{goulianos} and refs. therein.}. 
In both cases the constant rate indicates a deep connection between the
production mechanism of rapidity gap events and
the parton dynamics at small-$x$.

\begin{figure}
\vspace{-0.3cm}
\epsfxsize130pt
\mbox{\hspace{-0.3cm}
\figurebox{120}{120}{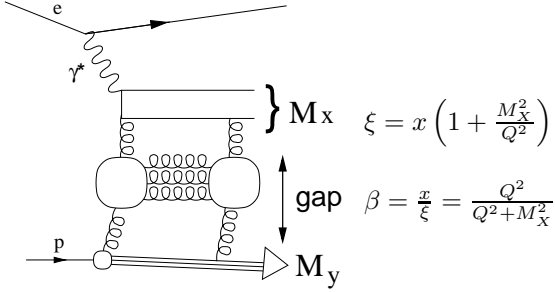}
}
\begin{picture}(50.,50.)
\put(0.,60) {$\Large \xi = x \left ( 1 + \frac{M_X^2}{Q^2} \right)$}
\put(0.,30) {$\Large \beta = \frac{x}{\xi} = \frac{Q^2}{Q^2 + M_X^2}$}
\end{picture}
\caption{Sketch of a diffractive scattering at HERA.}
\label{fig:diff}
\vspace{-0.3cm}
\end{figure}

\begin{figure}
\vspace{-0.5cm}
\epsfxsize210pt
\figurebox{120}{120}{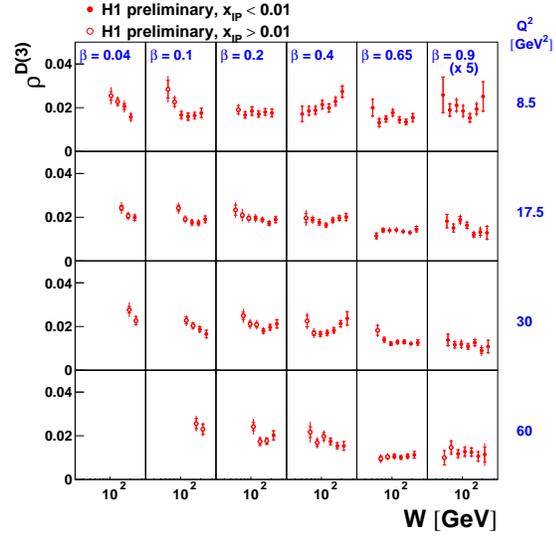}
\vspace{-1.cm}
\caption{Ratio of the diffractive to the inclusive DIS cross
section 
as function of the $\gamma^* p$ centre of
mass energy $W$ in bins of $Q^2$ and $\beta$.}
\label{fig:ratioh1}
\vspace{-0.6cm}
\end{figure}

The interaction responsible for a large rapidity gap 
can be viewed as the exchange of a colour singlet
decomposing the hadronic final state into a system $X$ and $Y$: 
$ep \to e X Y$ (see Fig.~\ref{fig:diff}).
The kinematic of the process can be described
by the longitudinal momentum fraction $\xi$ of the
colourless exchange with respect to the incoming proton\footnote{This quanity
is often called $x_{_{I\!\!P}}$.}
and, in analogy to Bjorken-$x$ in the inclusive case,
by the longitudinal momentum fraction $\beta$ that is
carried by the quark struck by the photon (see Fig.~\ref{fig:diff}).
%
%
%%%%%%%%%%%%%%%%%%%%%%%%%%%%%%%%%%%%%%%%%%%%%%%%%%
%%%%%%%%%%%%%%%%%%%%%%%%%%%%%%%%%%%%%%%%%%%%%%%%%%%%
The ratio\footnote{The diffractive cross section is in this analysis 
integrated over $M_Y < 1.6$ {\rm GeV} and $t < 1~{\rm GeV}^2$,
where $t$ is the squared momentum transfer from the proton to the system Y.} 
\begin{eqnarray}
\vspace{-0.2cm}
%                        1         2                  2         11       2                21
\rho^{D(3)} = M_X^2 \frac{d\sigma_{\gamma^* p \to X Y}/ dM_X^2}{\sigma_{\gamma^* p \to X}}
\nonumber
\vspace{-0.5cm}
\label{eq:seccdm}  
\end{eqnarray}
is shown in Fig.\ref{fig:ratioh1} as a function of the $\gamma^* p$ cm energy\cite{h1diffnew}. 
The ratio $\rho^{D(3)}$ is relatively flat over the full phase space.
Only at low $\beta$ a tendancy to rise is observed.
The constant rate as a function of the cm energy has already been observed 
in the first HERA data\cite{diffzeus}, but it is still rather surprising, when
one tries to understand its physical meaning.
In the simplest approach\footnote{
For full reviews of all discussed models see ref\cite{diffraction}.} 
the colour singlet exchange can be modelled by
the exchange of two ``gluons''\cite{twogluon}. If the ``gluons'' are
interpreted as partons, i.e. pQCD is applicable because a hard scale is
involved in the process, then at small-$x$ the inclusive cross section is
expected to rise like $\sigma \propto x^{-\lambda}$ while the diffractive
cross section should rise like $\sigma \propto x^{- 2 \lambda}$.
In this case the rapidity gap contribution would significantly increase
towards small-$x$. If the  colour singlet exchange is soft, the
increase of the diffractive cross section should be closely
related to soft hadron hadron and $\gamma p$
collisions where $\sigma \propto W^{0.08}$ and therefore
$\sigma \propto (1/x)^{0.16}$ is expected.
In this case the rapidity gap contribution in the DIS sample would decrease.

The constant behaviour of  $\rho^{D(3)}$ indicates that the production
mechanism of rapidity gap events contains soft and hard pieces.
It is remarkable that their interplay leads to the same energy 
dependence of the cross section as in the inclusive case.
%%%%%%%%%%%%%%%%%%%%%%%%%%%%%%%%%%%%%%%%%%%%%%%%%%
%%%%%%%%%%%%%%%%%%%%%%%%%%%%%%%%%%%%%%%%%%%%%%%%%%%%

For inclusive $\gamma^* p$ collisions, $Q^2$ provides the hard scale
to discriminate soft and hard processes. The energy behaviour of the
inclusive cross section (see eq.\ref{eq:secdis}) is shown in Fig.\ref{fig:lambda_zeus}.
Recently, ZEUS has achieved a remarkable precision on the inclusive cross
section measurements at very low $Q^2$ thanks to dedicated detectors close to the beam pipe
measuring electrons scattered at small angles\cite{ZEUSbpc}.
For $Q^2 \lsim 0.8~{\rm GeV}^2$, a constant slope $\lambda \approx 0.1$
is found. This value is consistent with the value ($\lambda = 0.08$) needed to describe
the total hadron hadron cross section at high energies\cite{dl}.
The same value also describes the energy behaviour of the
scattering of real photons on protons\cite{heragp}. A generalised vector
dominance model\cite{ZEUSvdm} describes the data.
For higher $Q^2$ values, $\lambda$ linearly rises with $\log{Q^2}$
which is well described by pQCD. The transition from the soft to the
hard regime seems to be surprisingly sharp ($0.5 \lsim Q^2 \lsim 1~{\rm GeV}^2$). 
Does this indicate that there are really two different physical
regimes, the long-distance, non-perturbative region, 
where only Regge theory can be applied, and  
the short-distance, pQCD region describing the
interaction of quarks and gluons ? Or is a unified
description based on small-$x$ phenomena possible ?

The physical understanding of the transition region
can be made more intuitive in the proton rest frame
rather than in the DIS frame, where the proton has infinite momentum.
In this frame the (virtual) photon splits into a $q\bar{q}$ pair
and develops a hadronic structure by radiating gluons
long before interacting with the proton. The timescale of this
fluctuation is $\tau \propto 1/ (x \, m_p)$.
At small-$x$, the lifetime $\tau$ of the hadronic system 
is long, 
e.g. at $x = 10^{-3}$, $c \tau = 200$~fm. 
The $q\bar{q}$ pair
can be viewed as a colour dipole
described by a wave function $\Psi_{T,L}$ depending on the
longitudinal and transverse photon polarisation\cite{nikolaev}.
It can be calculated within QED.
The $\gamma^* p$ cross section can be written as:
%%%%%%%%%%%%%%%%%%%%%%%%%%%%%%%%%%%%%%
\begin{eqnarray}
\vspace{-1.7cm}
\sigma(x,Q^2) = \int d^2r \, dz \; {|\Psi_{T,L}|}^2 \; \hat{\sigma}(x,r)  
% \nonumber
\vspace{-0.7cm}
\label{eq:secdiff}  
\end{eqnarray}
%%%%%%%%%%%%%%%%%%%%%%%%%%%%%%%%%%%%%%
where $r$ is the transverse separation of the $q\bar{q}$ dipole and
$z$ is the photon momentum fraction carried by the quark.
By the Heisenberg principle $r$ is closely related to the transverse
quark momentum via $r \propto 1/k_T$. 
For small $k_T$, i.e. large dipoles,
the colour field is large and it strongly interacts with the
proton. The dipole behaves as a hadron in soft collisions.
For large $k_T$, i.e. small dipoles, 
the colour field is effectively screened and the proton is
``transparent'' to the dipole (colour transparency). 
$\hat{\sigma}(x,r)$ describes the interaction 
of the $q\bar{q}$ dipole with the proton.

\begin{figure}
\vspace{-0.1cm}
\epsfxsize200pt
\figurebox{120}{120}{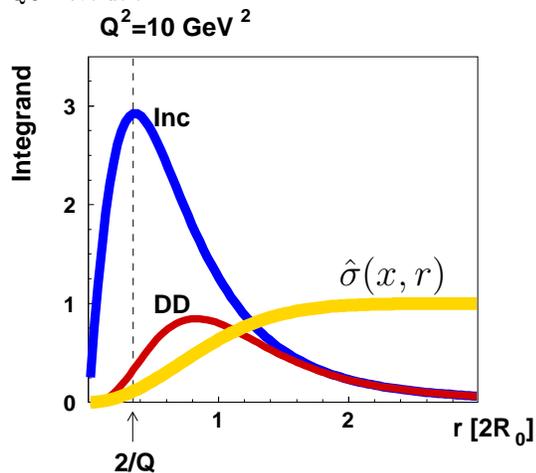}
\begin{picture}(50.,50.)
\put(125.,122.) {\Large $\hat{\sigma}(x,r)$}
\end{picture}
\vspace{-1.8cm}
\caption{The integrand in equation \ref{eq:seccdm} for
inclusive (Inc) and diffractive (DD) scattering 
and the dipole cross section $\hat{\sigma}(x,r)$.}
\label{fig:golec3}
\vspace{-0.7cm}
\end{figure}

The following phenomenological form\footnote{Another successful form inspired
by the generalised vector dominance model has been proposed in \cite{schildknecht}.} 
has been proposed\cite{golec} for $\hat{\sigma}(x,r)$:
%%%%%%%%%%%%%%%%%%%%%%%%%%%%%%%%%%%%%%
\begin{eqnarray}
\vspace{-1.8cm}
%\hat{\sigma(x,r)} = \sigma_0 \left 
\sigma_0 \left(1-\exp{\frac{- r^2}{4 R_0^2(x)}}\right) 
\, {\rm with} \,  
R_0^2(x)=\left(\frac{x}{x_0}\right)^{-\lambda} 
% \nonumber
\vspace{-0.8cm}
\label{eq:golec}  
\end{eqnarray}
%%%%%%%%%%%%%%%%%%%%%%%%%%%%%%%%%%%%%%  
where $\sigma_0 =23$~mb, $x_0 = 3 \cdot 10^{-4}$ and
$\lambda = 0.29$ are determined by a fit to inclusive DIS
data for $x < 0.01$\footnote{The radius $R_0$ has the units 1/{\rm GeV}.
These numbers are quoted for a model with no charm quarks.
Including charm quarks leads to larger values.}. 
As an example, the dipole cross section 
is shown at $Q^2 = 10~{\rm GeV}^2$ in Fig.~\ref{fig:golec3}.
For small $r$, the dipole cross section steeply rises 
$\hat{\sigma}(x,r) \propto r^2 x^{-\lambda}$ as expected by pQCD.
Towards large $r$ the cross section is assumed to saturate
$\hat{\sigma}(x,r) \propto \sigma_0$. It is interesting that the
integrand of equation \ref{eq:golec} shown in Fig.\ref{fig:golec3} 
peaks around $2/Q$ making the close relation between the photon virtuality
and the dipole size clear.

%%%%%%%%%%%%%%%%%%%%%%%%%%%%%%%%%%%%%%%%%%%%%%%%%%
\begin{figure}
%\vspace{-0.5cm}
%\epsfxsize175pt
%\figurebox{120}{120}{lambda.eps}
\epsfxsize155pt
%\mbox{\hspace{0.5cm}
\figurebox{120}{120}{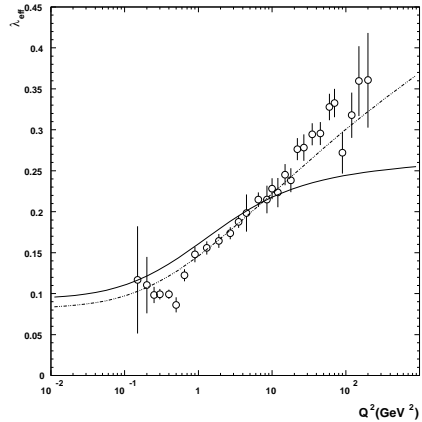}
%}
%\vspace{0.3cm}
%\begin{picture}(50.,50.)
%\put(15.,230.) {a)}
%\put(15., 50.) {b)}
%\end{picture}
%\vspace{-0.5cm}
\caption{Slope of the inclusive total $\gamma$p cross
section $\lambda$ as a function of $Q^2$. Shown as lines
are fits to the Golec-Biernat/W\"usthoff model with
and without QCD evolution. %, to a generalised vector dominance model
%and a QCD NLO DGLAP fit. 
}
\label{fig:lambda_golec}
\vspace{-0.6cm}
\end{figure}
%%%%%%%%%%%%%%%%%%%%%%%%%%%%%%%%%%%%%%%%%%%%%%%%%%%%

%%%%%%%%%%%%%%%%%%%%%%%%%%%%%%%%%%%%%%%%%%%%%%%%%%
%%%%%%%%%%%%%%%%%%%%%%%%%%%%%%%%%%%%%%%%%%%%%%%%%%%%
With the assumed form of the fitted dipole cross section 
the inclusive DIS data can be successfully described over a wide kinematic range
(see Fig.~\ref{fig:lambda_golec}b). The additional incorporation of
DGLAP evolution\cite{golec01} is needed to describe the data at large $Q^2$\footnote{
To do that, $1/R_0^2$ is replaced by \\
$4 \pi^2 \alpha_s(\mu^2) x g(x,\mu^2)/(3 \sigma_0)$ where
$\mu^2= C/r^2+\mu_0^2$, $x g(x,\mu^2) = A_g x^{-\lambda_g}$
and $C$, $A_g$, $\lambda_g$ and $\mu_0^2$ are additional fit
parameters.}. However, it can not fully account for the steep rise
of $F_2$ at large $Q^2$. 
The transition around
$Q^2 \approx 0.5~{\rm GeV}^2$ seems to be smoother than found in the
data. In view of problems to describe some of the recent diffractive data\cite{peppe}
it is questionable, if the DGLAP is the correct evolution to be applied
in this context. 

The diffractive cross section is obtained by modifying
equation \ref{eq:seccdm} for inclusive DIS\cite{nikolaev}:
%%%%%%%%%%%%%%%%%%%%%%%%%%%%%%%%%%%%%%
\begin{eqnarray}
\vspace{-0.3cm}
\sigma^{DD}(x,Q^2) = \int d^2r dz \; {|\Psi_{T,L}|}^2 \; \hat{\sigma}^2(x,r)
% \nonumber
\vspace{-0.2cm}
\label{eq:golec2}  
\end{eqnarray}
%%%%%%%%%%%%%%%%%%%%%%%%%%%%%%%%%%%%%%.
By analysing only the leading contribution a ratio
$\sigma^{DD}/\sigma^{inc} \propto 1/\ln{(Q^2 R_0^2(x)})$ is found
which is only slowly varying with $x$ and $Q^2$ and correctly
reproduces the data\cite{golecdiff} when in the dipole
also $q \bar{q} g$ states are included. 
Fig.~\ref{fig:golec3} shows that the integrand of eq.~\ref{eq:golec2}
contains a significantly higher contribution of large dipole sizes at the
same $Q^2$ than found in the inclusive case. 
This makes both the soft and the hard nature of the diffractive production mechanism
evident. By assuming a certain interaction of the colour dipole
with the proton based on the inclusive DIS data at small-x
the Golec-Biernat/W\"usthoff model predicts the diffractive cross section.
This underlines the deep connection between
diffraction and the small-$x$ parton dynamics which we start to 
understand better and better.
In the leading logarithm approximation the colour dipole formulation
and the $k_T$ factorisation
give an equivalent description of hard processes at high energy\cite{pesch}.

\vspace{-0.3cm}
\section{Instanton Production in DIS}
\label{sec:instanton}
\vspace{-0.3cm}
In QCD, anomalous non-perturbative processes are expected which
violate classical laws such as the conservation of chirality.
Instantons\cite{belavin}, non-perturbative fluctuations of the gluon field,
represent tunnelling transitions between topologically inequivalent vacua.
DIS offers a unique possibility to discover QCD instantons.
%%%%%%%%%%%%%%%%%%%%%%%%%%%%%%%%%%%%%%%%%%%%%%%%%%
\begin{figure}%1
\epsfxsize140pt
\vspace{-0.3cm}
\figurebox{120}{120}{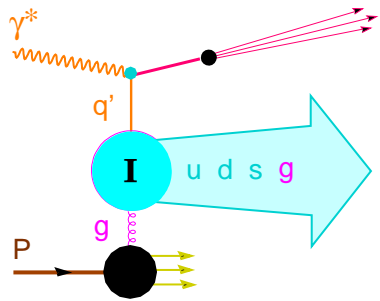}
%\vspace{0.3cm}
\caption{Sketch of the production of instanton induced
events in DIS.
}
\label{fig:inst_graph}
\vspace{-0.6cm}
\end{figure}
%%%%%%%%%%%%%%%%%%%%%%%%%%%%%%%%%%%%%%%%%%%%%%%%%%

In DIS instanton induced processes ($I$) are dominantly produced in
a quark gluon fusion process 
(see Fig.\ref{fig:inst_graph})~\cite{moch97,vladimir}. 
The virtuality $Q'^2$ of the quark $q'$,
originating from a photon splitting into a $q\bar{q}$ pair in
the $I$-background, provides a generic hard scale naturally limiting
the $I$-size $\rho$ and makes a quantitative prediction of the
cross-section possible~\cite{moch97,schrempp99}. 
The expected cross section for the kinematical
region $x > 10^{-3}$ and $0.1 < y < 0.9$ is of the order of
$10-100$ pb\cite{schrempp98,schrempp00}.
This cross section is sizeable, but still much lower than
the inclusive DIS cross section.
The expected characteristics of the hadronic final state 
of $I$-processes, 
which can be simulated using the QCDINS\cite{qcdins} program,
can be exploited to improve the signal to noise ratio\cite{gerigk}.
Besides the current jet,
a large number of particles of all kinematically allowed flavours
is produced at high $E_T$ 
in a narrow band of two rapidity units. 
Since the Instanton decays isotropically, a large sphericity 
of the particles in the $I$-rest system is expected.

\begin{figure}
%           height width name
\vspace{-0.2cm}
\begin{tabular}{cc}
\epsfxsize95pt
\figurebox{}{120}{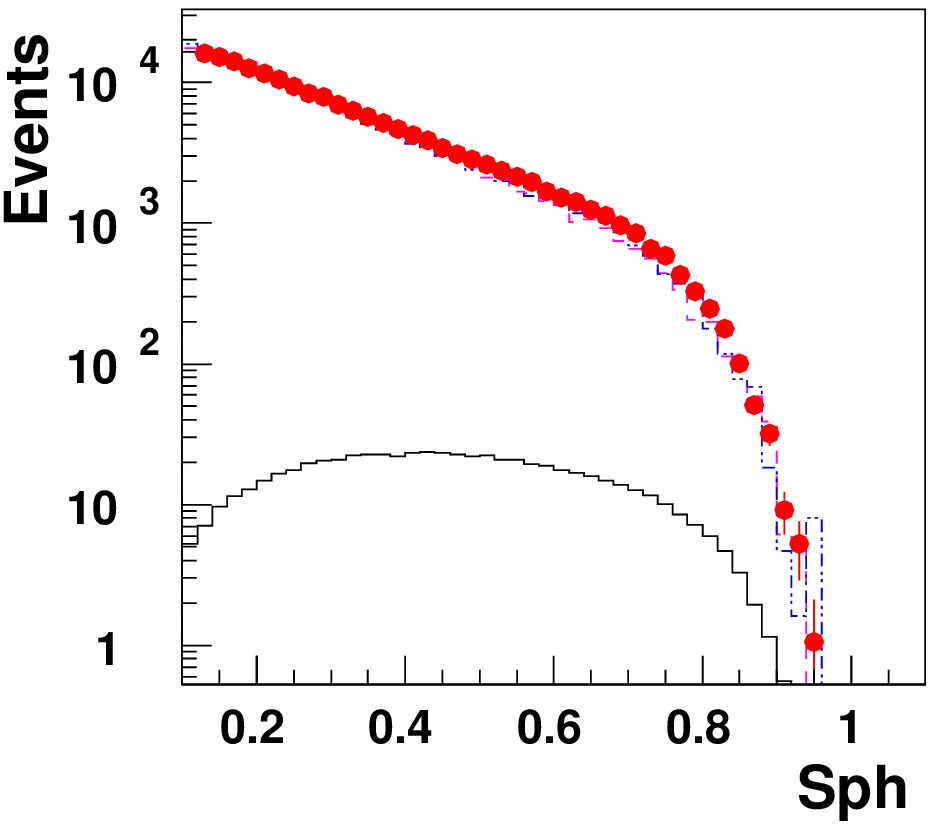}
\epsfxsize95pt
\figurebox{}{120}{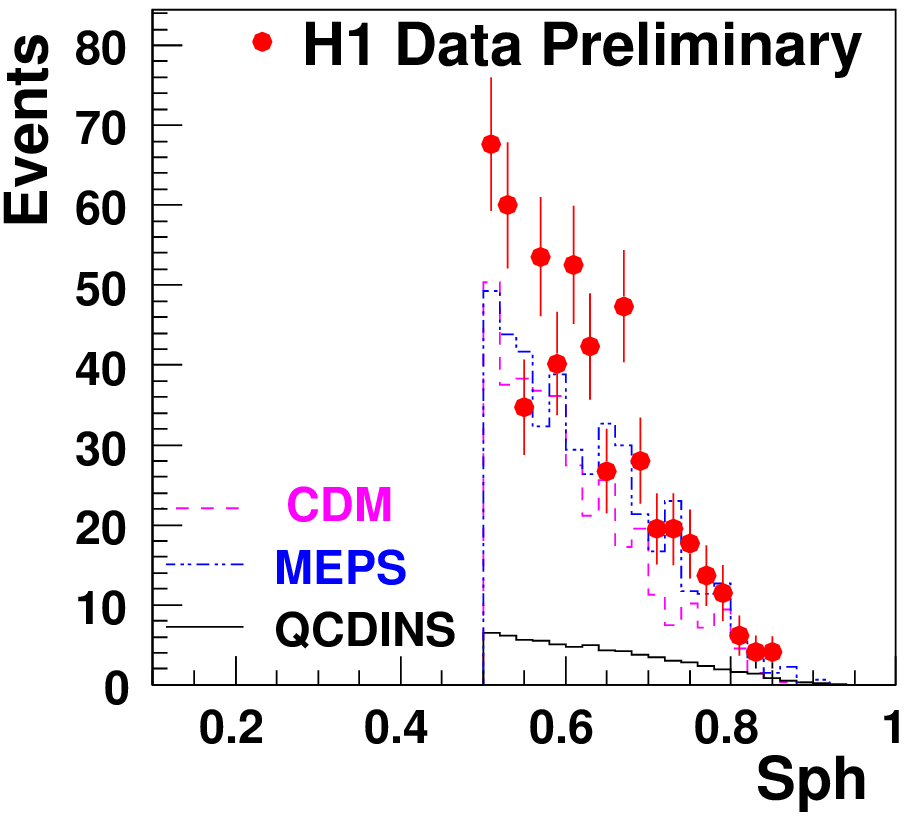}
\end{tabular}
\vspace{-1.6cm}
\begin{picture}(50.,50.)
\put(10.,55) {(a)}
\put(100.,55) {(b)}
\end{picture}
\vspace{-0.3cm}
\caption{Sphericity for normal and $I$-DIS
without (a) and with (b) cuts to enhance the $I$-signal. 
}
\label{fig:sphericity}
\vspace{-0.8cm}
\end{figure}

The H1 collaboration has recently performed\cite{h1instanton}
the first dedicated search for QCD $I$-induced processes 
using six hadronic final state $I$-sensitive observables in the kinematic range 
$x > 10^{-3}$, $0.1 < y < 0.6$ and $\theta_{el}>156^o$, where
$\theta_{el}$ is the polar angle of the scattered 
electron\footnote{For an $I$-analysis based on existing data 
see \cite{carliins}.}. 
As an example the sphericity distribution is shown 
in Fig.~\ref{fig:sphericity}a for the inclusive DIS sample.
Before cuts to enhance the $I$-fraction, the data are well described
by pQCD models. The expected instanton sample is about $2$-$3$ orders
of magnitude smaller and the events are expected to be more
spherical. After cuts the background is suppressed by about
a factor of $1000$. $549$ events are selected in the data, while only
$363^{+22}_{-26}$ and $435^{+36}_{-22}$ events are
expected by two different pQCD models. The size of the excess is, however,
about the same as the difference in the pQCD background models.
The sphericity distribution
of these events is shown in Fig.\ref{fig:sphericity}b.
In four out of the six $I$-sensitive observables a qualitative
similarity between the difference of the data and the pQCD models
and the expected $I$-shape is observed. The two remaining
observables do not particularly support the hypothesis that the
observed excess can be explained by I-processes as currently
implemented in QCDINS. However, recently it has been pointed out
that the shape of these two particular observables is affected by 
large theoretical uncertainties due to missing cuts in the
experimental analysis\cite{schremppzoom}.
The data limit the allowed $I$-cross section to about $100$~pb.
A steep rise of the I-cross section towards large instanton sizes,
as would be obtained from a naive extrapolation of $I$-perturbation
theory, is excluded by the data. The absence of such a steep rise
is in agreement with lattice simulations of zero flavours\cite{schrempp99,schrempp98}.

%%%%%%%%%%%%%%%%%%%%%%%%%%%%%%%%%%%%%%%%%%%%%%%%%%%%%%%%%%%
%%%%%%%%%%%%%%%%%%%%%%%%%%%%%%%%%%%%%%%%%%%%%%%%%%%%
%
\vspace{-0.5cm}
\section*{Conclusions}
\vspace{-0.2cm}
The strong cross section increase expected in
LO BFKL for various processes has not been observed. 
Thanks to a huge theoretical effort significant progress has been recently made.
The BFKL NLO  corrections have been calculated and additional
resummations have been worked out.
The results are in qualitative agreement with the data. However, no clear
and unmistakable evidence for the need for small-$x$ effects beyond
the standard resummation of leading logarithms using collinear factorisation
has been reported so far. 
Nevertheless there are strong indications that small-$x$ effects
play already at the present energies a certain role 
and that they describe high energy collisions in a more coherent way.
First encouraging results using angular ordering as a 
key unifying principle have been obtained.
This allows more precise QCD predictions for future colliders
to be made.

The increased precision of the HERA measurements reveals more and more the intriguing
connection between soft and hard physics. In the understanding of 
soft phenomena the correct description of diffraction plays a key role.
Here, the deep connection to small-$x$ parton dynamics becomes clearer and clearer
and more is to be learned in the near future. The final aim is to get 
better insights to the confinement of quark and gluons in hadrons.
The pioneering experimental and phenomenolgical work
to discover QCD instanton-induced processes might add important
complementary information in our understanding of non-perturbative 
phenomena. It seems possible to establish such an important non-perturbative
effect at HERA, although no clear experimental evidence has been found so far.
\vspace{-0.6cm}
\section*{Acknowledgements}
\vspace{-0.3cm}
It is a pleasure to thank my colleagues from the H1 and ZEUS
collaborations for providing me with the latest data and
J. Bartels, V. Chiochia,
C. Golec-Biernat, G. Iaccobucci, H. Jung, A. Martin, 
R. Nisius, 
G. Salam, F. Schrempp, R. Thorne   
and J. Whitmore
for invaluable help in preparing this talk
or for commenting on the manuscript.
%
%\vspace{-0.5cm}
\section*{Questions:}
\vspace{-0.4cm}
\subsection*{G. Altarelli, CERN:}
\vspace{-0.2cm} 
The beautiful HERA data on small-$x$ structure functions pose
a clear problem: Why do the NLO QCD evolution equations work so well
in spite of the fact that the BFKL corrections could be large.
You have mentioned the Salam, Ciafaloni, Colferai proposed
solution to this problem, but nobody has so far shown that
their theory fits the data well.
Myself, Ball and Forte we have developed an alternative
approach, less predictive, but more model independent
(in our opinion) which we have proven to fit the data
(we can accommodate but not predict the smallness of the
BFKL corrections). Until the understanding of totally
inclusive structure functions is not more satisfactory, the
theoretical foundations of many BFKL signals in
non inclusive distributions is quite shaky.
\vspace{-0.5cm}
\subsection*{Answer:}
\vspace{-0.2cm}
I agree that the $F_2$ data can be interpreted in the theoretically
cleanest way, since the uncertainties are usually much smaller
than for less inclusive observables. I have pointed out that
many open questions remain in the understanding of $F_2$ and
I have stressed the importance of even more precise data.
Since $F_2$ is a very inclusive quantity, precision is the key
issue. However, less inclusive observables allow to look
at new effects in a more direct way and therefore might be
more promising. At the end of the day, it will be important
to have a consistent theory of strong interactions to describe
inclusive and less inclusive measurements in many different
processes.
\vspace{-0.2cm}
\subsection*{P. Nason, INFN Milano:}
I have a remark with regard to the CCFM Monte Carlo predictions.
Monte Carlo programs are practical implementation of theoretical
ideas and calculations, and at the end they should match precise
analytical calculations. For example, attempts to explain the
beauty cross section by small-$x$ resummations, once 
${\cal O}(\alpha_s^2)$ and ${\cal O}(\alpha_s^3)$ corrections
are correctly implemented, failed (until now) to explain the
data. Thus, the inaccuracy of the Monte Carlo (that certainly
does not implement the  ${\cal O}(\alpha_s^3)$ correctly)
should be addressed before claims are made of solving
phenomenological problems.
\vspace{-0.5cm}
\subsection*{Answer:}
\vspace{-0.2cm}
The CASCADE MC is an implementation of the CCFM equation.
It seems to be adequate, since in
all proposed CCFM solutions the correct treatment of the kinematics
turned out to be important. 
Although it is based on LO, part of the higher orders
are resummed. In particular when $k_T$ factorisation is used, it is
difficult to say if, e.g. a ${\cal O}(\alpha_s^2)$ calculation based
on the collinear factorisation contains more or less higher order
contributions.
In this sense there is no inaccuracy, but it is a different approximation.
%
%\vspace{-0.2cm}
\subsection*{K. Ellis, Fermilab:}
You showed calculations due to Jung and Salam which lead to
a larger beauty quark cross section at the TEVATRON.
It is important to establish what feature of the calculations
give rise to the increase. Does the unintegrated gluon distribution
used in the calculations give (after integration over $k_T^2$)
a gluon distribution in accord with the gluon distribution measured
elsewhere at HERA ?
\vspace{-0.5cm}
\subsection*{Answer:}
\vspace{-0.2cm}
The gluon density is not a physical observable. Therefore the gluon
as extracted from NLO DGLAP fits and from the CCFM scheme do not have
to agree and in fact they do not. The unintegrated gluon is
higher \cite{jungring}. However,
it is only important that both describe the cross section
measurements and in fact they both do so. 
%
%
%%%%%%%%%%%%%%%%%%%%%%%%%%%%%%%%%%%%%%%%%%%%%%%%%%%%%%%%%%%%%%% 
%%%%%%%%%%%%%%%%%%%%%%%%%%%%%%%%%%%%%%%%%%%%%%%%%%%%%%%%%%%%%%%  

{\small
\begin{thebibliography}{99}
%
% F2charm ZEUS, EPJ C 12 (2000) 35
%
\bibitem{bfkl} % see C. Schmidt hep-ph/0106181 for more ref
E.A. Kuraev, L.N. Lipatov, V.S. Fadin, 
Sov. Phys. JETP 44 (1976) 443 and 
Sov. Phys. JETP 45 (1977) 199;
I.I. Balitsky, L.N. Lipatov, Sov. J. Nucl. Phys. 28 (1978) 822.
%
\bibitem{bfklnlo} % see G. Salam hep-ph/9910492 for more ref
V.S. Fadin, L.N. Lipatov, 
Phys. Lett. B 429 (1998) 127 % hep-ph/9802290
; see also: G. Camici, M. Ciafaloni, Phys. Lett B 412 (1997) 396
(erratum ibid. B 417 (1997) 390 % hep-ph/9707390
; G. Camici, M. Ciafaloni, Phys. Lett. B 430 (1998) 349.
%
\bibitem{bfklsol} % 
S.J. Brodsky et al., JETP Lett. 70 (1999); %hep-ph/9901229
J. Bl\"umlein, A. Vogt, Phys. Rev. D 57 (1998) 1; %hep-ph/9707488
ibid. D 58 (1998) 014020; %hep-ph/9912546
C. Schmidt, Phys. Rev. D 60 (1999) 074003; %hep-ph/9901397
J.R. Forshaw, D.A. Ross, A. Sabio Vera, 
Phys. Lett. B 455 (1999) 273; %hep-ph/9903390
J. Kwiecinski, A.D. Martin, P. Sutton,
Z. Phys. C 71 (1996) 585.
%
\bibitem{f2bfkl} J. Kwiecinski, A.D. Martin, A. Stasto,
Phys. Rev. D 56 (1997) 3991.
% 
\bibitem{bfklintro} % 
G. Salam, in: Cracow School of Theo. Phys. (1999) 
hep-ph/9910492;
C.R. Schmidt, in: 5th Int. Sym. on Rad. Corr., Carmel (USA),
hep-ph/0106181.
%
\bibitem{bfklresum} % 
M. Ciafaloni, D. Colferai, 
Phys. Lett. B 452 (1999) 372; % hep-ph/9812366
M. Ciafaloni, D. Colferai, G. Salam, 
Phys. Rev. D 60 (1999) 1140036; %hep-ph/9905566 
G. Salam, JHEP 07 (1998) 19. % hep-ph/9806482
%
\bibitem{dglap} % 
V.N. Gribov, L.N. Lipatov, Sov. J. Nucl. Phys. 15 (1972) 438;
G. Altarelli, G. Parisi, Nucl. Phys. B 126 (1977) 298;
Yu. L. Dokshitzer, Sov. Phys. JETP 46 (1977) 641.
%
\bibitem{martin93} A. Martin, Lect. XXI Int. Meet.
Fund. Phys., Madrid (1993) DTP/93/66.
%
\bibitem{rujula} %
A. De Rujula et al., Phys. Rev. D 10 (1974) 1649. 
%
\bibitem{ktfac} %
S. Catani, M. Ciafaloni, F. Hautmann, Nucl. Phys. B 366 (1991) 135;
J.C. Collins, R.K. Ellis, Nucl. Phys. B 360 (1991) 3; 
E.M. Levin et al., Sov. J. Nucl. Phys. 54 (1991) 867.
\bibitem{ccfm} %
M. Ciafaloni, Nucl. Phys. B 296 (1988) 49; 
S. Catani, F. Fiorani, G. Marchesini, Phys. Lett. B 234 (1990) 339
and Nucl. Phys. B 336 (1990) 18;
 G. Marchesini, Nucl. Phys. B 445 (1995) 49.
%
\bibitem{impact} %
J. Bartels, S. Gieseke, C.F. Qiao, Phys. Rev. D 63 (2001) 056014 %hep-ph/0009102 
and  hep-ph/0107152;
V.S. Fadin, D. Yu. Ivanov, M. I. Kotsky, hep-ph/0106099. 
%
\bibitem{kimber} M.A. Kimber, J. Kwiecinski, A.D. Martin, 
Phys. Lett. B 508 (2001) 58.
%
\bibitem{smallx} %
G. Marchesini, B. Webber, Nucl. Phys. B 349 (1991) 617
and Nucl. Phys. B 386 (1992) 215.
%
\bibitem{ccfmpheno} %
G. Bottazzi, G. Marchesini, G. Salam, M. Scorletti,
Nucl. Phys. B 505 (1997) 366 % hep-ph/9702418
and JHEP 9812 (1998) 011. % hep-ph/9810546
%
\bibitem{cascade} %
H. Jung, hep-ph/0109102, to be publ. Comp. Phys. Com.
%
\bibitem{ccfmhadfin} %
H. Jung, G. Salam, Eur. Phys. J. C 19 (2001) 351. %hep-ph/0012143
%
\bibitem{ldc} %
B.Andersson et al.,
Nucl. Phys. B 467 (1996) 443; Z. Phys. C 71 (1996) 613;
Nucl. Phys. B 467 (1996) 443;
H. Kharrazhia, L. L\"onnblad, JHEP 03 (1998) 6. %hep-ph/9709424
%
%\bibitem{jungdis00} %
%H. Jung and L. L\"onnblad, J. Phys. G. Nucl. Phys. 26 (2000) 707.
%
%%% pp jets at large rapidity
\bibitem{mueller87} %
A.H. Mueller, H. Navelet, Nucl. Phys. B 282 (1987) 727.
%
\bibitem{d0bfkl} %
D0 coll., Phys. Rev. Lett. 84 (2000) 5722.
%
\bibitem{orr98} %
L.H. Orr, W. J. Stirling, Phys. Lett. B 429 (1998) 127.
%
\bibitem{delduca01} %
J.R. Andersen et al., JHEP 02 (2001) 007. % hep-ph/0101180
%
\bibitem{symcuts} %
S. Frixione, G. Ridolfi, Nucl. Phys. B 507 (1997) 315; %hep-ph/9707345 
Kramer, Klasen, Phys. Lett. B 366 (1996) 385;
H1 coll., Eur. Phys. J. C 13 (2000) 415;
T. Carli, hep-ph/9906541.
%
%%% gamma/gamma
%
%\bibitem{phojet} %
%R. Engel, Z. Phys. C 66 (1995) 203; R. Engel and J. Ranft,
%Phys. Rev. D 54 (1996) 4246.
%
\bibitem{bartelgamgam} %
J. Bartels, A. de Roeck, H. Lotter, Phys. Lett. B 389 (1996) 742;
J. Bartels, C. Ewerz, R. Staritzbichler, Phys. Lett. B 492 (2000) 56.
%
\bibitem{kim99} %
S.J. Brodsky et al., JETP Lett. 70 (1999) 155;
and V. Kim private comm.
%
\bibitem{l3} %
L3 coll., Phys. Lett. B 453 (1999) 333.
%
\bibitem{l3_01} %
L3 coll., L3 note 2680, LP01.
\bibitem{cacciari01} 
M. Cacciari et al., JHEP 0102 (2001) 029. %hep-ph/0011368
%
\bibitem{opal} 
OPAL coll., hep-ex/0110006 
subm. to Eur. Phys. J. C. 
%
\bibitem{motyka} %
J. Kwiecinski, L. Motyka, Phys. Lett. B 462 (1999) 203.
%
% F2
%
 \bibitem{F2}
 H1 coll., 
 Nucl. Phys. B 407 (1993) 515
 and B 439 (1995) 471;
 Nucl. Phys. B 470 (1996) 3;
 Eur. Phys. J. C 13 (2000) 609.
 ZEUS coll., 
 Phys. Lett. B 316 (1993) 412;
 Z. Phys. C 65 (1995) 379;
 and C 69 (1996) 607;
%
 \bibitem{F2new} H1 coll., 
Eur. Phys. J. C 19 (2001) 269;
and C 21 (2001) 33;
ZEUS coll., Eur. Phys. J. C 21 (2001) 443.
%
\bibitem{DGLAPfit} M. Botje, Eur. Phys. J. C 10 (2000) 285;
V. Barone, C. Pascaud, F. Zomer, hep-ph/0004268;
ZEUS coll., contr. P1-628 to EPS01. 
%
 \bibitem{mrst} A.D. Martin et al.,
Eur. Phys. J. C 4 (1998) 463 and  C 14 (2000) 13.
%
\bibitem{cteq}
CTEQ coll., Eur. Phys. J. C 12 (2000) 375.
%
\bibitem{h1f2slope} 
H1 coll., Phys. Lett. B 520 (2001) 183.
%
\bibitem{thorne} R. Thorne, Phys. Lett. B 474 (2000) 372
and Phys. Rev. D 64 (2001) 074005. 
%
%
\bibitem{alta01} G. Altarelli, R.D. Ball, S. Forte, 
Nucl. Phys. B 599 (2001) 383.
%
 \bibitem{mrst01} A.D. Martin et al., hep-ph/0110215.
%
 \bibitem{d0jet} D0 coll., Phys. Rev. Lett. 86 (2001) 1707.
%
 \bibitem{cdfjet} CDF coll., Phys. Rev. D 64 (2001) 032001.
%
\bibitem{thorne2} R. Thorne, talk given at the Weimar QCD network 
meeting, 2001.
%
% heavy quark
%
%
\bibitem{h1charm} H1 coll., hep-ex/0108039. 
% 
\bibitem{hvqdis} B.W. Harris, J. Smith, 
Nucl. Phys. B 452 (1995) 109, % hep-ph/9503484
Phys. Lett. B 353 (1995) 535, % hep-ph/9502312
Phys. Rev. D 57 (1998) 2806. % hep-ph/9706334
%
%\bibitem{riemers} E. Laenen et al.,
%Nucl. Phys. B 392 (1995) 162,%
%S. Riemersma et al.,
%Phys. Lett. B 347 (1995) 143. % hep-ph/9411431
%
\bibitem{peterson} C. Peterson et  al.,
Phys. Rev. D 27 (1983) 105.
%
\bibitem{zeuscharm} ZEUS coll.,
Eur. Phys. J. C 12 (2000) 35. % hep-ex/9908012
%
\bibitem{norbin} E. Norrbin, T. Sj\"ostrand,
Eur. Phys. J. C 17 (2000) 137. % hep-ph/0005110
%
\bibitem{h1beautydis} H1 collab,
contr: 807, LP01
%
\bibitem{h1beauty}
H1 coll., Phys. Lett. B 467 (1999) 156;
erratum ibid B 518 (2001) 331. 
%
\bibitem{zeusbeauty}
ZEUS coll., Eur. Phys. J. C 18 (2001) 625. % hep-ex/0011081
%
\bibitem{jungring} H. Jung, hep-ph/0109146.
%
\bibitem{beautylep} L3 coll., Phys. Lett. B 503 (2001) 10; % hep-ex/0011070
OPAL note PN455.
%
\bibitem{tevatronbeauty} CDF coll., Phys. Rev. D 55 (1997) 2546;
D0 coll., Phys. Lett. B 487 (2000) 264; %hep-ph/9605270
and Phys. Rev. Lett. 85 (2000) 5068. % hep-ex/0008021
%
\bibitem{tevatronnlo} M. L. Mangano, P. Nason, G. Ridolfi, 
Nucl. Phys. B 373 (1992) 295. 
%
\bibitem{frixione97} S. Frixione, M.L. Mangano, 
Nucl. Phys. B 483 (1997) 321. % hep-ph/9605270
%
\bibitem{junghq} H. Jung, hep-ph/0110034 
%
\bibitem{hagler} 
P. H\"agler et al., Phys. Rev. D 62 (2000) 071502
%
% forward jet
%
\bibitem{fjets} A.H. Mueller, Nucl. Phys. B (Proc. Suppl.) 18 C (1990) 125;
 J. Phys. G 17 (1991) 1443;
 J. Bartels, A. de Roeck, M. Loewe, Z. Phys. C 54 (1992) 635.
% 
\bibitem{fjetdata} ZEUS collab, 
Phys. Lett. B 474 (2000) 1 and 
Eur. Phys. J. C 6 (1999) 239;
H1 coll., Nucl. Phys. B 538 (1999) 3.
%
%\bibitem{mirkesfjet} E. Mirkes and D. Zeppenfeld, 
%Phys. Rev. Lett. 78 (1997) 428. % hep-ph/9609231
%
\bibitem{fjetbfkl} J. Bartels et al., 
Phys. Lett. B 384 (1996) 300.
%
\bibitem{bfklfwdjet} % 
J. Kwiecinski, A. Martin, J. Outhwaite, Eur. Phys. J. C 9 (1999) 611.
%
\bibitem{jungresolved} % 
H. Jung, L. J\"onsson, H. K\"uster, Eur. Phys. J C 9 (1999) 383.
%
%
\bibitem{kramer99} %
G. Kramer, B. P\"otter, Phys. Lett. B 453 (1999) 295.
%
% \bibitem{lepto} G. Ingelman, in Proc. HERA workshop,
% Eds. W. B\"uchm\"uller and G. Ingelman, Hamburg (1991) Vol. 3,
% 1366; G. Ingelman, A. Edin and J. Rathsman, DESY-96057,
% Hamburg, April 1996. 
%
%\bibitem{partonshower} M. Bengtsson and T. Sj\"ostrand,
% Z. Phys. C37 (1998) 465; M. Bengtsson, G. Ingelman and
% T. Sj\"ostrand, Proc. of the HERA workshop 1987, ed. R.D. Peccei,
% DESY, Hamburg  Vol. 1 (1988) 149.
%
% \bibitem{lund} B. Andersson et al., Phys. Rep. 97 (1983) 31.
%
% \bibitem{ariadne} L. L\"onnblad, Computer Phys. Comm. 71 (1992) 15.
%
% \bibitem{cdm} G. Gustafson and U. Peterson, Nucl. Phys. B306 (1988);
% G. Gustafson, L. L\"onnblad and U. Peterson, Z. Phys. C43 (1989) 625. 
%
% \bibitem{rat96} J. Rathsman, 
% Differences between Monte Carlo models for DIS at small-$x$
%and the relation to BFKL dynamics
% SLAC-PUB-7344, October 1996.
%
\bibitem{schoerner} T. Schoerner, DIS 2000.
%
 \bibitem{kuhlen} M. Kuhlen, Phys. Lett. B 382 (1996) 441.
%
\bibitem{fwdpizero} H1 coll., Phys. Lett. B 462 (1999) 440.
%
% diffraction
%
\bibitem{diffhera} ZEUS coll., Phys. Lett. B 315 (1993) 481 and
B 332 (1994) 228; Z. Phys. C 68 (1995) 569; Eur. Phys. J. C 1 (1998) 81
and C 6 (1999) 43; 
H1 coll., Nucl. Phys. B 429 (1994) 477; 
Phys. Lett. B 348 (1995) 681; Nucl. Phys. B 472 (1996) 3;
Z. Phys. C 69 (1995) 27 and C 75 (1997) 607 and C 76 (1997) 613. 
%
\bibitem{goulianos} K. Goulianos, hep-ex/0011060,
J. Bartels and H. Kowalski, hep-ph/0010345.
%
\bibitem{h1diffnew} H1 coll., contr: 200, LP01
%
\bibitem{diffzeus} ZEUS coll., Phys. Lett. B 315 (1993) 481.
%
\bibitem{diffraction} M. W\"usthoff, A.D. Martin, J. Phys. G 25 (1999) R309,
%hep-ph/9909362,
M.F. McDermott, hep-ph/0008260; 
A. Hebecker,  Acta Phys. Polon. B 30 (199) 3777 and
Phys. Rep. 331 (2000) 1.
%
\bibitem{twogluon} F.E. Low, Phys. Rev. D 12 (1975) 163;
S. Nussinov, Phys. Rev. 34 (1976) 1286;
J. Gunion, D.E. Soper, Phys. Rev. D 15 (1977) 2617;
E.M. Levin, M.G. Ryskin, Sov. J. Nucl. Phys. 34 (1981) 619.
%
\bibitem{dl} A. Donnachie, P.V. Landshoff, 
Phys. Lett. B 296 (1992) 227.
%
%
\bibitem{ZEUSbpc} ZEUS coll., 
Phys. Lett. B 487 (2000) 1 and EPS01.
%
\bibitem{ZEUSvdm} ZEUS coll., Phys. Rev. D 53 (2000) 53.
%
\bibitem{heragp} H1 coll., Z. Phys. C 69 (1995) 27;
ZEUS coll.,  Z. Phys. C 63 (1994) 391
and contr: 1046, ICHEP 2000.
%
\bibitem{nikolaev} N. Nikolaev, B.G. Zakharov, Z. Phys. C 49 (1990) 607
and  Z. Phys. C 53 (1992) 331;
A.H. Mueller, Nucl. Phys. B 415 (1994) 373 and
Nucl. Phys. B 437 (1995) 107;
%
\bibitem{pesch} A. Bialas, H. Navelet, R. Peschanski, 
 Nucl. Phys. B 593 (2001) 438. %  hep-ph/0009248
%
\bibitem{golec} K. Golec-Biernat, M. W\"usthoff, 
Phys. Rev. D 59 (1999) 014017. 
%
\bibitem{schildknecht} D. Schildknecht et al., Phys. Lett. B 499 (2001) 116;
G. Cvetic et al., Eur. Phys. J. C 13 (2000) 301 and C 20 (2001) 77.
%
\bibitem{golecdiff} K. Golec-Biernat, M. W\"usthoff, 
Phys. Rev. D 60 (1999) 114023. 
%
\bibitem{golec01} K. Golec-Biernat, hep-ph/0109010.
%
\bibitem{peppe} G. Iaccobucci, these proceedings.
%
% Instantons
%
%\bibitem{braun}
%I.I. Balitskii, V.M. Braun,
% Phys. Lett. B 346 (1995) 143.
%
\bibitem{belavin}
A. Belavin et al., Phys. Lett. B 59 (1975) 85;
G. t'Hooft, Phys. Rev. D 14 (1976) 3432 and  
Phys. Rev. Lett. 37 (1976) 8.
%
\bibitem{moch97}  S. Moch, A. Ringwald, F. Schrempp,
 Nucl. Phys. B 507 (1997) 134.
%
\bibitem{vladimir}  A. Ringwald,F. Schrempp,hep-ph/9411217.
%
\bibitem{schrempp99}
A. Ringwald, F. Schrempp, Phys. Lett. B 459 (1999) 249.
%
\bibitem{schrempp98}
A. Ringwald, F. Schrempp, Phys. Lett. B 438 (1998) 217.
%
\bibitem{schrempp00}
A. Ringwald,F. Schrempp,hep-ph/9909338, \\hep-ph/0006215.
%
\bibitem{qcdins}
A. Ringwald, F. Schrempp, Comp. Phys. Commun. 132 (2000) 267. 
%
%
\bibitem{gerigk} T. Carli et al., hep-ph/9906441.
%
\bibitem{h1instanton} H1 coll.,
contr. paper to ICHEP00.
%
\bibitem{carliins}
T. Carli, M. Kuhlen, Nucl. Phys. B 511 (1998) 85. 
%
\bibitem{schremppzoom}
A. Ringwald, F. Schrempp, Phys. Lett. B 503 (2001) 331.
%
\end{thebibliography}
}
\end{document}